\documentclass[english,journal,letterpaper]{IEEEtran}
\usepackage[T1]{fontenc}
\usepackage[latin9]{inputenc}
\usepackage{amsmath}
\usepackage{amssymb}
\usepackage{graphicx}

\makeatletter
\usepackage{cite}
\usepackage{times}   
\newtheorem{theorem}{Theorem}[section]
\newtheorem{definition}[theorem]{Definition}
\newtheorem{remark}[theorem]{Remark}
\newtheorem{corollary}[theorem]{Corollary}
\newtheorem{newexample}[theorem]{Example}
\newtheorem{lemma}[theorem]{Lemma}
\newtheorem{proposition}[theorem]{Proposition}
\newtheorem{aplemma}{Lemma}[section]

\usepackage[all]{xy}

\usepackage{balance}

\usepackage{tikz}
\usetikzlibrary{positioning, arrows, matrix}

\usepackage{psfrag}

\usepackage{dsfont}

\newtheorem{assumption}{Assumption}[section]

\usepackage{url}

\mathchardef\mhyphen="2D

\renewcommand{\IEEEQED}{\IEEEQEDopen}

\makeatother

\usepackage{babel}
\begin{document}

\title{\LARGE \bf Networked Control under Random and Malicious Packet Losses}

\author{Ahmet Cetinkaya, Hideaki Ishii, and Tomohisa Hayakawa \thanks{A. Cetinkaya and T. Hayakawa are with the Department of Mechanical and Environmental Informatics, Tokyo Institute of Technology, Tokyo 152-8552, Japan. {\tt\small{ahmet@dsl.mei.titech.ac.jp, hayakawa@mei.titech.ac.jp}}}
\thanks{H. Ishii is with the Department of Computational Intelligence and Systems Science, Tokyo Insitute of Technology, Yokohama, 226-8502, Japan. {\tt\small{ishii@dis.titech.ac.jp}}}
\thanks{This work was supported in part by Japan Science and Technology Agency under the EMS-CREST program.}}

\maketitle
\begin{abstract} We study cyber security issues in networked control
of a linear dynamical system. Specifically, the dynamical system and
the controller are assumed to be connected through a communication
channel that face malicious attacks as well as random packet losses
due to unreliability of transmissions. We provide a probabilistic
characterization for the link failures which allows us to study combined
effects of malicious and random packet losses. We first investigate
almost sure stabilization under an event-triggered control law, where
we utilize Lyapunov-like functions to characterize the triggering
times at which the plant and the controller attempt to exchange state
and control data over the network. We then provide a look at the networked
control problem from the attacker's perspective and explore malicious
attacks that cause instability. Finally, we demonstrate the efficacy
of our results with numerical examples. \end{abstract}

\section{Introduction}

Cyber security has become a critical problem in industrial processes,
since nowadays they incorporate information and communication technologies
that are prone to cyber threats. Cyber attacks can disrupt the normal
operation of services that are critical to the society as they can
cause financial losses and environmental damages. It is thus essential
to ensure cyber security of existing infrastructures and design new
cyber-attack-resilient ones. 

Literature on cyber security points out cyber threats against industrial
control systems utilized in many fields (see \cite{cardenas2008research}
and the references therein). Vulnerabilities of the channels used
for transmission of measurement and control data pose a critical issue
for the security of control systems. This is because the channels
are recently connected via the Internet or wireless communications
\cite{fawzi2014,wholejournal2015}. Communication channels, for instance,
may face jamming attacks initiated by malicious agents \cite{xu2005feasibility,pelechrinis2011}.
Such attacks block the communication link and effectively prevent
transmission of packets between the plant and the controller. It is
mentioned in \cite{pelechrinis2011} that jamming attacks pose a major
security threat, as they can be easily performed with devices that
target various wireless communication protocols. In recent works \cite{amin2009,lee2013modeling,bhattacharya2013,HSF-SM:ct-13,depersis2014,de2015inputtran,liu2014stochastic,li2015jamming},
networked control problems under jamming attacks were investigated
using control and/or game-theoretic methods. However, jamming may
not be the only cause of malicious packet losses. Compromised routers
in a network may also intentionally drop packets \cite{mizrak2009detecting,shu2015privacy}.
The work \cite{d2013fault} explored the control problem over a multihop
network with malicious nodes that intentionally stop forwarding packets
or alter packet contents. 

In addition to actions of malicious agents, state measurement and
control input packets may also fail to be transmitted at times due
to network congestion or errors in communication. Stochastic models
provide accurate characterization of such nonmalicious network issues
\cite{khayam2003markov,altman2005}. In the literature, unreliability
of a network is often characterized through random models for packet
loss events \cite{schenato2007,hespanha2007}. For instance, in \cite{Xiong2007,ishii2009,Lemmon:2011:ASS:1967701.1967744},
Bernoulli processes are used for modeling packet losses in a network.
Furthermore, in \cite{gupta2009,okano2014}, packet loss events are
characterized in a more general way by employing Markov chains. In
those studies, a variety of control methods are also proposed to ensure
stability of networked control systems that face random packet losses. 

In this paper, we propose a stochastic representation of packet transmission
failures in a network between a plant and a controller. Our proposed
model is sufficiently general and allows us to explore some of the
existing random and malicious packet loss scenarios in a unified manner.
At the core of this characterization, we have a tail probability condition
on the average number of state measurement and control input packet
failures in the network. We demonstrate that random packet losses,
malicious attacks, as well as the combination of those two phenomena
satisfy the condition with different parameters. We model random losses
by using a binary-valued \emph{time-inhomogeneous} Markov chain. Furthermore,
to characterize malicious attacks, we use a model similar to the one
in \cite{depersis2014}. Specifically, this model allows attacks to
happen arbitrarily as long as the total number of packet exchange
attempts that face malicious attacks are almost surely bounded by
a certain ratio of the number of total packet exchange attempts between
the plant and the controller. The almost sure bound used in our model
in fact allows not only deterministic strategies but also stochasticity
in the generation of malicious attacks. As a result, the model captures
attacks that are generated based on randomly varying information such
as state and control input or the random packet losses. Besides, an
attacker may also intentionally use randomness to imitate packet losses
that occur due to congestion or channel noise. 

Through our malicious attack model, we consider scenarios where the
attacker targets the network only when the plant and the controller
attempt to exchange packets. In a jamming attack scenario, our characterization,
hence, can be considered as a model for \emph{reactive} jamming discussed
in \cite{xu2005feasibility} for wireless networks. The classification
in \cite{xu2005feasibility} divides attackers into two groups: \emph{active}
and \emph{reactive} ones. An \emph{active} jamming attacker tries
to block a communication channel regardless of whether the channel
is being used or not, whereas a \emph{reactive} attacker continuously
monitors the channel and attacks only when there is transmission.
It is mentioned in \cite{xu2005feasibility} that it may be harder
to detect a \emph{reactive} jamming attacker as packets may also be
lost due to nonmalicious network issues and hence the reason for packet
losses may not be known with certainty. A similar issue where packet
losses occur due to both malicious and nonmalicious reasons exists
also in the context of multihop networks. For instance \cite{shu2015privacy}
investigates combined effects of malicious packet drops and nonmalicious
channel errors. 

Motivated by the scenarios mentioned above, we utilize our probabilistic
characterization also to investigate networks that are subject to
the combination of random transmission errors due to unreliability
of the channel and attacks conducted by malicious agents. In our analysis,
we consider two cases: (i) when the attacks and random packet losses
are modeled as independent processes and (ii) when the attack strategy
is dependent on the random packet losses. The dependent case is essential
to model the situation where the attacker has information of the random
packet losses in the communication channel and utilizes this information
in the attack strategy. Furthermore, we may also consider situations
when the attacker decides to attack based on the content of packets.
In the case of jamming attacks, this corresponds to selective jamming
discussed in \cite{proano2010selective,proano2012packet}, where the
intelligent jamming attacker listens to the communication channel
and decides whether to interfere or not depending on the packet being
transmitted. For example, a jamming attacker may decide not to interfere
with the communication when the packet being transmitted is already
corrupted by channel noise. Moreover, in a network of multiple nodes
malicious ones may intentionally drop certain packets based on their
content \cite{xiao2007chemas}. The main theoretical challenge in
dealing with the combination of random packet losses and malicious
attacks stems from the fact that these two phenomena are of different
nature and hence have different models. By utilizing a tail probability
inequality for the sum of processes that represent random packet losses
and malicious attacks, we show that our proposed probabilistic characterization
allows us to deal with both independent and dependent loss cases. 

By utilizing our probabilistic packet transmission model, we investigate
the networked control problem of a linear plant through an event-triggered
framework. Event-triggered control methods have recently been employed
in many studies (see \cite{tabuada2007event,heemels2012cdc,liu2014survey}
and the references therein). We follow the approach in \cite{velasco2009,heemels2013periodic}
and utilize Lyapunov-like functions to determine the triggering times
at which the plant and the controller attempt to exchange state and
control input information. The triggering conditions that we propose
ensure that the value of a Lyapunov-like function of the state stays
within certain limits. Packet exchanges are attempted only before
the value of the Lyapunov-like function is predicted to exceed the
limit. In a successful packet exchange scenario, state measurements
are sent from the plant to the controller, which computes a control
input and sends it back to the plant. However, state measurement or
control input packets may fail to be transmitted due to random packet
losses and malicious attacks. 

Our packet failure characterization and control system analysis differ
from those of the recent studies \cite{qu2015event,rabi2009scheduling,quevedo2014stochastic},
which also investigate the event-triggered control problem under packet
losses. Specifically, in \cite{qu2015event}, the number of consecutive
packet losses is assumed to be upper-bounded, and a deterministic
Lyapunov function approach is used for the closed-loop stability analysis.
Moreover, in \cite{rabi2009scheduling,quevedo2014stochastic} the
packet losses are modeled by a Bernoulli process. The stability analysis
in \cite{quevedo2014stochastic} is based on investigating the evolution
of the expectation of a Lyapunov function. Despite the similarity
to our malicious attack model, our stability analysis also differs
from that of \cite{depersis2014}, where the analysis relies on a
deterministic approach for obtaining an exponentially decreasing upper
bound for the norm of the state. Our approach for stability analysis
is related to obtaining an upper bound on the \emph{top Lyapunov exponent}
(see \cite{fang1995stability,ezzine1997almost,bolzern2004almost})
of the system and in that sense it is more similar to the stability
analysis conducted in \cite{kellett2005stability,Lemmon:2011:ASS:1967701.1967744}
for networked systems without event-triggering. Specifically, we find
a stochastic upper bound for a Lyapunov-like function and show that
this stochastic upper bound tends to zero under certain conditions
indicating \emph{almost sure asymptotic stability}. 

In addition to stability analysis, we also address the question of
finding \emph{instability} conditions under which the state of the
closed-loop system diverges almost surely. We observe that an attack
strategy that causes sufficiently frequent packet losses can destabilize
the closed-loop dynamics. This instability result allows us to investigate
effects of potential malicious attacks on a networked control system. 

The rest of the paper is organized as follows. In Section~\ref{sec:Network-Control-Problem},
we describe the networked control problem under random and malicious
packet losses. We present an event-triggered control framework and
provide sufficient conditions for almost sure asymptotic stability
of the closed-loop system in Section~\ref{sec:Event-Triggered-Control-Design}.
In Section~IV, we look at the networked problem from the attacker's
perspective and provide conditions for instability of the system.
We present illustrative numerical examples in Section~\ref{sec:Numerical-Example}.
Finally, in Section~VI, we conclude the paper. 

We note that part of the results in Sections~\ref{sec:Network-Control-Problem}
and \ref{sec:Event-Triggered-Control-Design} appeared without proofs
in our preliminary report \cite{ahmetcdc2015}. Here, we provide a
more detailed discussion with complete proofs. 

We use a fairly standard notation in the paper. Specifically, we denote
positive and nonnegative integers by $\mathbb{N}$ and $\mathbb{N}_{0}$,
respectively. Moreover, $\|\cdot\|$ denotes the Euclidean vector
norm and $\left\lfloor \cdot\right\rfloor $ denotes the largest integer
that is less than or equal to its real argument. The notation $\mathrm{\mathbb{P}}[\cdot]$
denotes the probability on a probability space $(\Omega,\mathcal{F},\mathbb{P})$
with filtration $\{\mathcal{F}_{i}\}_{i\in\mathbb{N}_{0}}$ such that
$\mathcal{F}_{i_{1}}\subset\mathcal{F}_{i_{2}}\subset\mathcal{F}$
for $i_{1},i_{2}\in\mathbb{N}_{0}$ with $i_{1}<i_{2}$.

\section{Networked Control Problem and Characterization of Network with Random
and Malicious Packet Losses \label{sec:Network-Control-Problem}}

In this section we introduce the networked control problem and present
a characterization for a network with random packet losses and those
caused by malicious agents.

\subsection{Networked Control System}

Consider the linear dynamical system 
\begin{align}
x(t+1) & =Ax(t)+Bu(t),\quad x(0)=x_{0},\quad t\in\mathbb{N}_{0},\label{eq:system}
\end{align}
 where $x(t)\in\mathbb{R}^{n}$ and $u(t)\in\mathbb{R}^{m}$ denote
the state and the control input, respectively; furthermore, $A\in\mathbb{R}^{n\times n}$
and $B^{n\times m}$ are the state and input matrices, respectively. 

In our networked control problem, the plant and the controller exchange
information packets over a communication channel to achieve stabilization
of the zero solution $x(t)\equiv0$. We consider the case where packets
are transmitted without delay, but they may get lost. In a successful
packet exchange scenario, at a certain time instant, measured plant
states are transmitted to the controller, which generates a control
input signal and sends it to the plant. The transmitted control input
is applied at the plant side. In the case of an unsuccessful packet
exchange attempt, either the measured state packet or the control
input packet may get dropped, and in such cases control input at the
plant side is set to $0$, which is a common approach in the literature
(e.g., \cite{kellett2005stability,hespanha2007,gupta2009,okano2014}).
In this setup, the plant is informed about a packet exchange failure
by the lack of an incoming control input. Specific acknowledgement
messages are thus not needed. This allows the practical implementation
by using a UDP-like communication protocol discussed in \cite{schenato2007}. 

We use $\tau_{i}\in\mathbb{N}_{0},i\in\mathbb{N}_{0}$, (with $\tau_{i}<\tau_{i+1}$)
to denote the time instants at which packet exchanges between the
plant and the controller are attempted. In this paper, we consider
both the case where packet exchanges are attempted at all time instants
and the case where an event-triggering mechanism decides the successive
packet exchange attempt times. In both cases, the control input $u(t)$
applied to the plant is given by 
\begin{align}
u(t) & \triangleq\left(1-l(i)\right)Kx(\tau_{i}),\,t\in\{\tau_{i},\ldots,\tau_{i+1}-1\},\label{eq:control-input}
\end{align}
where $K\in\mathbb{R}^{m\times n}$ denotes the feedback gain and
$\{l(i)\in\{0,1\}\}_{i\in\mathbb{N}_{0}}$ is a binary-valued process
that characterizes success or failure of packet exchange attempts.
When $l(i)=0$, the packet exchange attempt at time $\tau_{i}$ is
successful and the piecewise-constant control input at the plant side
is set to $u(\tau_{i})=Kx(\tau_{i})$. On the other hand, $l(i)=1$
indicates that either the packet sent from the plant or the packet
sent from the controller is lost at time $\tau_{i}$. Again, in such
situations, control input at the plant side is set to $0$. We emphasize
that the framework described above allows us to deal with dropouts
in both state and control input channels of the network illustrated
in Fig.~\ref{operation}. In particular, the process $l(\cdot)$
is an \emph{overall} indicator of the packet exchange failures over
these channels.

\subsection{Network Characterization}

Packet transmission failures in a network may have different reasons.
In what follows we characterize the effects of certain stochastic
and malicious packet loss models in a unified manner by exploring
dynamical evolution of the total number of packet exchange failures. 

First, we define a nonnegative integer-valued process $\{L(k)\in\mathbb{N}_{0}\}_{k\in\mathbb{N}}$
by 
\begin{align}
L(k) & \triangleq\sum_{i=0}^{k-1}l(i),\quad k\in\mathbb{N}.\label{eq:bigldefinition}
\end{align}
 Note that $L(k)$ denotes the total number of \emph{failed} packet
exchange attempts during the time interval $[0,\tau_{k-1}]$, where
$k$ attempts have been made.

In our packet loss model, we place a bound on the ratio of failed
attempts in a probabilistic and asymptotic sense. 

\begin{assumption} \label{MainAssumption} There exists a scalar
$\rho\in[0,1]$ such that 
\begin{align}
\sum_{k=1}^{\infty}\mathbb{P}[L(k)>\rho k] & <\infty.\label{eq:lcond1}
\end{align}

\end{assumption} 

The condition (\ref{eq:lcond1}) provides a probabilistic characterization
of the evolution of the total number of packet exchange failures through
the scalar $\rho\in[0,1]$, representing their average ratio. Note
also that (\ref{eq:lcond1}) describes a condition on the tail probability
$\mathbb{P}[L(k)>\rho k]=\mathbb{P}[\frac{L(k)}{k}>\rho]$ of loss
ratio $\frac{L(k)}{k}$. This condition is sufficiently general and
includes some of the existing packet loss models in the literature.
We illustrate its generality by establishing that condition (\ref{eq:lcond1})
holds for four different cases: 
\begin{enumerate}
\item random packet losses,
\item malicious packet losses,
\item combination of the two losses in 1) and 2) when they are independent,
and finally
\item combination but when they are dependent.
\end{enumerate}
Note that for any packet loss model, Assumption~\ref{MainAssumption}
is trivially satisfied with $\rho=1$, since $\mathbb{P}[L(k)>k]=0$.
On the other hand, as we see below, for certain random and malicious
packet loss models, $\rho$ can be obtained to be strictly smaller
than $1$. A closely related characterization for packet dropouts
is presented in \cite{Lemmon:2011:ASS:1967701.1967744}; the scalar
$\rho$ in (\ref{eq:lcond1}) corresponds to the notion of \emph{dropout
rate} discussed there.

\subsubsection{Random Packet Losses}

\label{RemarkRandomPacketLoss}

To characterize nonmalicious network issues such as packet drops due
to network congestion or communication errors, we utilize time-inhomogeneous
Markov chains. Specifically, let $\{l_{\mathrm{R}}(i)\in\{0,1\}\}_{i\in\mathbb{N}_{0}}$
be a time-inhomogeneous Markov chain adapted to filtration $\{\mathcal{F}_{i}\}_{i\in\mathbb{N}_{0}}$.
Here, the $\sigma$-algebra $\mathcal{F}_{i}$ contains all random
packet transmission success/failure events for the first $i+1$ packet
exchange attempt times $\{\tau_{0},\tau_{1},\ldots,\tau_{i}\}.$ The
Markov chain $\{l_{\mathrm{R}}(i)\in\{0,1\}\}_{i\in\mathbb{N}_{0}}$
is characterized by initial distributions $\vartheta_{q}\in[0,1]$,
$q\in\{0,1\}$, and time-varying transition probabilities $p_{q,r}\colon\mathbb{N}_{0}\to[0,1]$,
$q,r\in\{0,1\}$, such that 
\begin{align}
\begin{array}{c}
\mathbb{P}[l_{\mathrm{R}}(0)=q]=\vartheta_{q},\\
\mathbb{P}[l_{\mathrm{R}}(i+1)=r|l_{\mathrm{R}}(i)=q]=p_{q,r}(i),\quad i\in\mathbb{N}_{0}.
\end{array}\label{eq:markov-chain-def}
\end{align}
The state $l_{\mathrm{R}}(i)=1$ indicates that the network faces
random packet losses at time $\tau_{i}$, and hence the packet exchange
attempt at $\tau_{i}$ results in failure. Here, success/failure of
a packet exchange attempt depends on the states of the previous packet
exchange attempts. Furthermore, transition probabilities between success
($l_{\mathrm{R}}(i)=0$) and failure ($l_{\mathrm{R}}(i)=1$) states
are time-dependent. It is important to note that the time-inhomogeneous
Markov chain characterization with time-varying transition probabilities
allows us to take into account the variation in the network between
consecutive packet transmission instants. Furthermore, this characterization
generalizes the Bernoulli and \emph{time-homogeneous} Markov chain
models that are often used in the literature.

In what follows we show that Assumption~\ref{MainAssumption} is
satisfied when the network faces random packet losses described by
\emph{time-inhomogeneous }Markov chains. In characterization of the
scalar $\rho$ used in Assumption~\ref{MainAssumption} we use upper-bounds
for transmission failure and success probabilities denoted respectively
by $p_{1}\in[0,1]$ and $p_{0}\in[0,1]$ such that 
\begin{align}
p_{q,1}(i) & \leq p_{1},\label{eq:p1condition-1}\\
p_{q,0}(i) & \leq p_{0},\quad q\in\{0,1\},\quad i\in\mathbb{N}_{0}.\label{eq:p0condition-1}
\end{align}
Note that even though $p_{q,r}(i)$ provide precise information about
the transitions between the states of random packet losses, this information
cannot be utilized when the network faces the combination of malicious
attacks and random packet losses (discussed in Sections~\ref{RemarkCombinedIndependent}
and \ref{RemarkCombinedDependent}). In such cases, information about
the probability of malicious attacks for each transmission attempt
is not available, and as a result, transition probabilities $p_{q,r}(i)$
for random packet losses cannot be utilized to obtain the overall
packet exchange failure probabilities. On the other hand, we can employ
the upper-bounds $p_{1}$ and $p_{0}$ when we show that the overall
packet exchange failures satisfy Assumption~\ref{MainAssumption}. 

\begin{lemma} \label{Lemma-Random-Loss} For the time-inhomogeneous
process $\{l_{\mathrm{R}}(i)\in\{0,1\}\}_{i\in\mathbb{N}_{0}}$ with
transmission failure probability upper-bound $p_{1}\in(0,1)$ that
satisfy (\ref{eq:p1condition-1}), we have 
\begin{align}
\sum_{k=1}^{\infty}\mathbb{P}[\sum_{i=0}^{k-1}l_{\mathrm{R}}(i)>\rho_{\mathrm{R}}k] & <\infty,\label{eq:random-loss-satisfies-assumption}
\end{align}
for all $\rho_{\mathrm{R}}\in(p_{1},1)$. 

\end{lemma}

\begin{IEEEproof} We use Lemma~\ref{KeyMarkovLemma} in the Appendix
to prove this result. Specifically, let $\tilde{p}=p_{1}$, $\tilde{w}=1$,
and define the processes $\{\xi(i)\in\{0,1\}\}_{i\in\mathbb{N}_{0}}$
and $\{\chi(i)\in\{0,1\}\}_{i\in\mathbb{N}_{0}}$ with $\xi(i)=l_{\mathrm{R}}(i)$
and $\chi(i)=1$, $i\in\mathbb{N}_{0}$. Since the conditions in (\ref{eq:xicond})
and (\ref{eq:chicond}) are satisfied, it follows from Lemma~\ref{KeyMarkovLemma}
that $\mathbb{P}[\sum_{i=0}^{k-1}l_{\mathrm{R}}(i)>\rho_{\mathrm{R}}k]\leq\psi_{k}$,
where $\psi_{k}\triangleq\phi^{-\rho_{\mathrm{R}}k+1}\frac{\left((\phi_{1}-1)p_{1}+1\right)^{k}-1}{(\phi-1)p_{1}}$
with $\phi\triangleq\frac{\rho_{\mathrm{R}}(1-p_{1})}{p_{1}(1-\rho_{\mathrm{R}})}$,
and $\sum_{k=1}^{\infty}\psi_{k}<\infty$, which implies (\ref{eq:random-loss-satisfies-assumption}).
\end{IEEEproof}

Lemma~\ref{Lemma-Random-Loss} indicates that when packet exchange
failures occur due to random packet losses (i.e., $l(i)=l_{\mathrm{R}}(i)$),
Assumption~\ref{MainAssumption} holds for all $\rho\in(p_{1},1)$.

\subsubsection{Packet Losses Due to Malicious Activity}

\label{RemarkJammingAttacks} 

Packet transmissions in a channel may get interrupted due to malicious
activities. For example, a compromised router in a network may deny
to forward incoming packets. In addition, packet losses may also be
caused by jamming attacks. A model for the attack strategy of a malicious
agent has been proposed in \cite{depersis2014}. In that study, the
sum of the length of attack durations is assumed to be bounded by
a certain ratio of total time.

By following the approach of \cite{depersis2014}, let $\{l_{\mathrm{M}}(i)\in\{0,1\}\}_{i\in\mathbb{N}_{0}}$
denote the state of attacks. The state $l_{\mathrm{M}}(i)=1$ indicates
that the packet transmission faces an attack at time $\tau_{i}$.
We consider the case where the number of packet exchange attempts
that face attacks are upper bounded almost surely by a certain ratio
of the total number of packet exchange attempts, that is, $\{l_{\mathrm{M}}(i)\in\{0,1\}\}_{i\in\mathbb{N}_{0}}$
satisfies 
\begin{align}
\mathbb{P}\big[\sum_{i=0}^{k-1}l_{\mathrm{M}}(i)\leq\kappa+\frac{k}{\tau}\big]=1,\quad k\in\mathbb{N},\label{eq:jammingcondition}
\end{align}
where $\kappa\ge0$ and $\tau>1$. In this characterization, among
$k$ packet exchange attempts, at most $\kappa+\frac{k}{\tau}$ of
them are affected by attacks. Note that when $\kappa=0$, (\ref{eq:jammingcondition})
implies no attack in the beginning: $l_{\mathrm{M}}(i)=0$, $i\in\{0,\ldots,\lfloor\tau\rfloor\}$,
almost surely. Scenarios that involve possible attacks during the
first few packet exchange attempts can be modeled by setting $\kappa>0$. 

In what follows, we would like to highlight the relations of the malicious
packet loss model in (\ref{eq:jammingcondition}) to those in the
literature. First, since the attacks only happen at packet exchange
attempt instants, the characterization in (\ref{eq:jammingcondition})
can be considered as a \emph{reactive jamming} model \cite{xu2005feasibility},
where the attacker attacks the  channel only when there is a packet
being transmitted. To avoid being detected, an attacker may refrain
from causing all packets to be lost. The ratio $\frac{1}{\tau}$ in
(\ref{eq:jammingcondition}) characterizes the average portion of
the packet transmission attempts that face attacks. Furthermore, in
the case of jamming attacks, in addition to avoid being detected,
the attacker may also need to take into account the energy requirements
of jamming. The ratio $\frac{1}{\tau}$ in this case corresponds to
the notion \emph{jamming rate} discussed in \cite{anantharamu2011},
and it is related to the energy usage of the jammer. 

\begin{remark}A packet loss model that may be used to capture behavior
of an intelligent attacker is also discussed in \cite{Xiong2007},
where transmissions between the plant and the controller are attempted
at all time instants and the proposed model allows packet losses to
occur arbitrarily as long as the lengths of intervals between consecutive
successful packet transmissions are not more than a given fixed length.
A similar model has also been used in \cite{qu2015event}, where an
event-triggered control method is used and the number of consecutive
packet losses is assumed to be upper-bounded by a constant. Note that
the packet loss model discussed in \cite{Xiong2007,qu2015event} can
be described within the framework provided by (\ref{eq:jammingcondition})
through setting $\tau=\frac{s+1}{s}$, where $s\geq1$ denotes the
upper-bound on the number of consecutive packet losses. Under this
setting, the condition (\ref{eq:jammingcondition}) provides more
freedom to the attacker as it does not necessarily require lengths
of intervals between consecutive successful packet transmission times
to be upper-bounded by a fixed constant. In fact for any $\tau>1$,
(\ref{eq:jammingcondition}) allows the attacker to cause any number
of consecutive packet losses after waiting sufficiently long without
attacking. Notice that the number of consecutive packet losses is
not restricted to be bounded also in the case of random packet loss
 models (see Section~\ref{RemarkRandomPacketLoss}, as well as \cite{hespanha2007,gupta2009,quevedo2012}).
\end{remark}

As pointed out in \cite{depersis2014}, the condition (\ref{eq:jammingcondition})
also shares some similarities with the socalled average dwell time
condition \cite{hespanha1999stability} utilized in switched systems.
In switched systems, the average dwell time condition requires the
number $N(k_{2},k_{1})$ of switches in between times $k_{1}$ and
$k_{2}\geq k_{1}$ to satisfy 
\begin{align}
N(k_{2},k_{1}) & \leq\kappa+\frac{k_{2}-k_{1}}{\tau},\quad k_{2}\geq k_{1}\geq0,\label{eq:average-dwell-time-condition}
\end{align}
 where $\tau>0$ denotes the average dwell time. The inequality (\ref{eq:average-dwell-time-condition})
guarantees that the switches occur slowly on average. In this study,
we do not require a condition on the number of switches between packet
exchange success and failure states. Rather than that we utilize (\ref{eq:jammingcondition}),
which is a condition on the total number of packet exchange failures
due to attacks. The condition (\ref{eq:jammingcondition}) guarantees
that attacks happen rarely on average. Note also that when $N(k_{2},k_{1})$
is defined to denote the number of packet exchange failures due to
attacks over all packet exchange attempts at times $\tau_{k_{1}},\tau_{k_{1}+1},\ldots,\tau_{k_{2}-1}$,
(\ref{eq:average-dwell-time-condition}) implies (\ref{eq:jammingcondition}).
Specifically, (\ref{eq:average-dwell-time-condition}) reduces to
(\ref{eq:jammingcondition}) by setting $N(k_{2},k_{1})\triangleq\sum_{i=k_{1}}^{k_{2}-1}l_{\mathrm{M}}(i)$,
$k_{1}=0$, and $k_{2}=k$. 

As we have observed so far, the attack model in (\ref{eq:jammingcondition})
is sufficiently general to cover known models. We further generalize
it, because even though the model in (\ref{eq:jammingcondition})
allows stochasticity in the generation of $l_{\mathrm{M}}(\cdot)$,
it is not enough to characterize certain stochastic attacks. An example
is the case where each packet exchange attempt faces an attack with
a fixed probability (e.g., $\{l_{\mathrm{M}}(i)\in\{0,1\}\}_{i\in\mathbb{N}_{0}}$
is a Bernoulli process). To cover such stochastic attacks as well
as attacks characterized in (\ref{eq:jammingcondition}), we consider
a model where $\{l_{\mathrm{M}}(i)\in\{0,1\}\}_{i\in\mathbb{N}_{0}}$
is given through conditions similar to (\ref{eq:lcond1}). Specifically,
we assume that there exists a scalar $\rho_{\mathrm{M}}\in[0,1]$
such that 
\begin{align}
\sum_{k=1}^{\infty}\mathbb{P}[\sum_{i=0}^{k-1}l_{\mathrm{M}}(i)>\rho_{\mathrm{M}}k] & <\infty.\label{eq:general-attack-condition}
\end{align}
The following lemma shows that the characterization with (\ref{eq:general-attack-condition})
is more general than the one provided by (\ref{eq:jammingcondition}). 

\begin{lemma} \label{JammingLemma} Suppose the binary-valued process
$\{l_{\mathrm{M}}(i)\in\{0,1\}\}_{i\in\mathbb{N}_{0}}$ satisfies
(\ref{eq:jammingcondition}) with $\kappa\ge0$ and $\tau>1$. Then
(\ref{eq:general-attack-condition}) holds for all $\rho_{\mathrm{M}}\in(\frac{1}{\tau},1)$. 

\end{lemma}

\begin{IEEEproof}Using Markov's inequality we obtain 
\begin{align}
 & \mathbb{P}[\sum_{i=0}^{k-1}l_{\mathrm{M}}(i)>\rho_{\mathrm{M}}k]\leq\mathbb{P}[\sum_{i=0}^{k-1}l_{\mathrm{M}}(i)\geq\rho_{\mathrm{M}}k]\nonumber \\
 & \quad=\mathbb{P}[e^{\sum_{i=0}^{k-1}l_{\mathrm{M}}(i)}\geq e^{\rho_{\mathrm{M}}k}]\leq e^{-\rho_{\mathrm{M}}k}\mathbb{E}[e^{\sum_{i=0}^{k-1}l_{\mathrm{M}}(i)}]\label{eq:markovsineqforattack}
\end{align}
for $k\in\mathbb{N}$. By (\ref{eq:jammingcondition}), we have $\mathbb{E}[e^{\sum_{i=0}^{k-1}l_{\mathrm{M}}(i)}]\leq\mathbb{E}[e^{\kappa+\frac{k}{\tau}}]=e^{\kappa+\frac{k}{\tau}}$.
Therefore, it follows from (\ref{eq:markovsineqforattack}) that $\mathbb{P}[\sum_{i=0}^{k-1}l_{\mathrm{M}}(i)>\rho_{\mathrm{M}}k]\leq e^{\kappa-(\rho_{\mathrm{M}}-\frac{1}{\tau})k},$
$k\in\mathbb{N}$. Thus, for all $\rho_{\mathrm{M}}\in(\frac{1}{\tau},1)$,
\begin{align*}
 & \sum_{k=1}^{\infty}\mathbb{P}[\sum_{i=0}^{k-1}l_{\mathrm{M}}(i)>\rho_{\mathrm{M}}k]\leq\sum_{k=1}^{\infty}e^{\kappa-(\rho_{\mathrm{M}}-\frac{1}{\tau})k}\\
 & \quad=e^{\kappa}e^{-(\rho_{\mathrm{M}}-\frac{1}{\tau})}\left(1-e^{-(\rho_{\mathrm{M}}-\frac{1}{\tau})}\right)^{-1}<\infty,
\end{align*}
which completes the proof. \end{IEEEproof}

Thus, if the only cause of packet losses is attacks (i.e., $l(i)=l_{\mathrm{M}}(i)$),
then Assumption~\ref{MainAssumption} holds with $\rho=\rho_{\mathrm{M}}$.

\subsubsection{Combination of Random and Malicious Packet Losses (independent case)}

\label{RemarkCombinedIndependent} 

In order to model the case where the network is subject to both random
and malicious packet losses, we define $\{l(i)\in\{0,1\}\}_{i\in\mathbb{N}_{0}}$
by 
\begin{align}
l(i) & =\begin{cases}
1,\quad & l_{\mathrm{R}}(i)=1\,\,\mathrm{or}\,\,l_{\mathrm{M}}(i)=1,\\
0,\quad & \mathrm{otherwise},
\end{cases}\,\,\,i\in\mathbb{N}_{0},\label{eq:ldefinitionforcombinedcase}
\end{align}
 where $\{l_{\mathrm{R}}(i)\in\{0,1\}\}_{i\in\mathbb{N}_{0}}$ is
a time-inhomogeneous Markov chain given in (\ref{eq:markov-chain-def})
characterizing random packet losses (from Section~\ref{RemarkRandomPacketLoss})
and $\{l_{\mathrm{M}}(i)\in\{0,1\}\}_{i\in\mathbb{N}_{0}}$ satisfying
(\ref{eq:general-attack-condition}) is a binary-valued process that
represents attacks of a malicious agent (from Section~\ref{RemarkJammingAttacks}). 

Proposition~\ref{PropositionCombinedCase} below provides a range
of values for $\rho\in(0,1)$ that satisfy Assumption~\ref{MainAssumption}
in the case where the network faces both random and malicious packet
losses. 

\begin{proposition} \label{PropositionCombinedCase} Consider the
packet exchange failure indicator process $\{l(i)\in\{0,1\}\}_{i\in\mathbb{N}_{0}}$
given by (\ref{eq:ldefinitionforcombinedcase}) where $\{l_{\mathrm{R}}(i)\in\{0,1\}\}_{i\in\mathbb{N}_{0}}$
and $\{l_{\mathrm{M}}(i)\in\{0,1\}\}_{i\in\mathbb{N}_{0}}$ are mutually
independent. Assume 
\begin{align}
p_{1}+p_{0}\rho_{\mathrm{M}} & <1,\label{eq:propositioncondition1}
\end{align}
where $p_{1},p_{0}\in(0,1)$ are scalars that satisfy (\ref{eq:p1condition-1}),
(\ref{eq:p0condition-1}). Then (\ref{eq:lcond1}) holds for all $\rho\in(p_{1}+p_{0}\rho_{\mathrm{M}},1)$.

\end{proposition} 

\begin{IEEEproof} From (\ref{eq:ldefinitionforcombinedcase}), the
overall loss process can be given by 
\begin{align*}
l(i) & =l_{\mathrm{R}}(i)+(1-l_{\mathrm{R}}(i))l_{\mathrm{M}}(i),\quad i\in\mathbb{N}_{0},
\end{align*}
 and hence, by (\ref{eq:bigldefinition}), 
\begin{align}
L(k) & =\sum_{i=0}^{k-1}l_{\mathrm{R}}(i)+\sum_{i=0}^{k-1}(1-l_{\mathrm{R}}(i))l_{\mathrm{M}}(i),\quad k\in\mathbb{N}.\label{eq:Linusefulform}
\end{align}
Now, let $\epsilon\triangleq\rho-p_{1}-p_{0}\rho_{\mathrm{M}}$, $\epsilon_{2}\triangleq\min\{\frac{\epsilon}{2},\frac{\rho_{\mathrm{M}}-p_{0}\rho_{\mathrm{M}}}{2}\}$,
$\epsilon_{1}\triangleq\epsilon-\epsilon_{2}$, and define $\rho_{1}\triangleq p_{1}+\epsilon_{1}$,
$\rho_{2}\triangleq p_{0}\rho_{\mathrm{M}}+\epsilon_{2}$. Furthermore,
let $L_{1}(k)\triangleq\sum_{i=0}^{k-1}l_{\mathrm{R}}(i)$ and $L_{2}(k)\triangleq\sum_{i=0}^{k-1}(1-l_{\mathrm{R}}(i))l_{\mathrm{M}}(i)$.
We then have 
\begin{align}
\mathbb{P}[L(k)>\rho k] & =\mathbb{P}[L_{1}(k)+L_{2}(k)>\rho_{1}k+\rho_{2}k]\nonumber \\
 & \leq\mathbb{P}[\left\{ L_{1}(k)>\rho_{1}k\right\} \cup\left\{ L_{2}(k)>\rho_{2}k\right\} ]\nonumber \\
 & \leq\mathbb{P}[L_{1}(k)>\rho_{1}k]+\mathbb{P}[L_{2}(k)>\rho_{2}k].\label{eq:keyprobabilityinequality}
\end{align}
In the following we will show that the series $\sum_{k=1}^{\infty}\mathbb{P}[L_{1}(k)>\rho_{1}k]$
and $\sum_{k=1}^{\infty}\mathbb{P}[L_{2}(k)>\rho_{2}k]$ are convergent. 

First, note that 
\begin{align}
\rho_{1} & =p_{1}+\epsilon-\epsilon_{2}=\max\{p_{1}+\frac{\epsilon}{2},p_{1}+\epsilon-\frac{\rho_{\mathrm{M}}-p_{0}\rho_{\mathrm{M}}}{2}\}\nonumber \\
 & =\max\{\frac{p_{1}+\rho-p_{0}\rho_{\mathrm{M}}}{2},\rho-p_{0}\rho_{\mathrm{M}}-\frac{\rho_{\mathrm{M}}-p_{0}\rho_{\mathrm{M}}}{2}\}\nonumber \\
 & =\max\{\frac{p_{1}+\rho-p_{0}\rho_{\mathrm{M}}}{2},\frac{2\rho-\rho_{\mathrm{M}}(1+p_{0})}{2}\}.\label{eq:rho1ineq}
\end{align}
As $\frac{p_{1}+\rho-p_{0}\rho_{\mathrm{M}}}{2}<1$ and $\frac{2\rho-\rho_{\mathrm{M}}(1+p_{0})}{2}<1$,
it holds from (\ref{eq:rho1ineq}) that $\rho_{1}\in(p_{1},1)$. Consequently,
$\sum_{k=1}^{\infty}\mathbb{P}[L_{1}(k)>\rho_{1}k]<\infty$ follows
from Lemma~\ref{Lemma-Random-Loss} with $\rho_{\mathrm{R}}$ replaced
with $\rho_{1}$. 

Next, we will use Lemma~\ref{KeyMarkovLemma} to show that $\sum_{k=1}^{\infty}\mathbb{P}[L_{2}(k)>\rho_{2}k]<\infty$.
To obtain this result, we first observe that $\rho_{2}>p_{0}\rho_{\mathrm{M}}$,
since $\epsilon_{2}>0$. Moreover, 
\begin{align*}
\rho_{2} & =p_{0}\rho_{\mathrm{M}}+\min\{\frac{\epsilon}{2},\frac{\rho_{\mathrm{M}}-p_{0}\rho_{\mathrm{M}}}{2}\}\leq p_{0}\rho_{\mathrm{M}}+\frac{\rho_{\mathrm{M}}-p_{0}\rho_{\mathrm{M}}}{2}\\
 & <p_{0}\rho_{\mathrm{M}}+\rho_{\mathrm{M}}-p_{0}\rho_{\mathrm{M}}=\rho_{\mathrm{M}},
\end{align*}
and hence, we have $\rho_{2}\in(p_{0}\rho_{\mathrm{M}},\rho_{\mathrm{M}})$.
As a consequence of (\ref{eq:jammingcondition}), conditions (\ref{eq:xicond}),
(\ref{eq:chicond}) in the Lemma~\ref{KeyMarkovLemma} hold with
$\tilde{p}=p_{0}$ and $\tilde{w}=\rho_{\mathrm{M}}$, together with
processes $\{\xi(i)\in\{0,1\}\}_{i\in\mathbb{N}_{0}}$ and $\{\chi(i)\in\{0,1\}\}_{i\in\mathbb{N}_{0}}$
defined by setting $\xi(i)=1-l_{\mathrm{R}}(i)$, $\chi(i)=l_{\mathrm{M}}(i)$,
$i\in\mathbb{N}_{0}$. Now, we have $L_{2}(k)=\sum_{i=0}^{k-1}\xi(i)\chi(i)$
and hence, Lemma~\ref{KeyMarkovLemma} implies $\sum_{k=1}^{\infty}\mathbb{P}[L_{2}(k)>\rho_{2}k]<\infty$. 

Finally, by (\ref{eq:keyprobabilityinequality}), we arrive at 
\begin{align*}
 & \sum_{k=1}^{\infty}\mathbb{P}[L(k)>\rho k]\\
 & \quad\leq\sum_{k=1}^{\infty}\mathbb{P}[L_{1}(k)>\rho_{1}k]+\sum_{k=1}^{\infty}\mathbb{P}[L_{2}(k)>\rho_{2}k]<\infty,
\end{align*}
which completes the proof. \end{IEEEproof}

\subsubsection{Combination of Random and Malicious Packet Losses (dependent case)}

\label{RemarkCombinedDependent} 

So far, in Proposition~\ref{PropositionCombinedCase}, we assumed
that packet exchange attempt failures due to attacks are independent
of those due to random packet losses. Next, we consider the case where
the two processes $\{l_{\mathrm{R}}(i)\}_{i\in\mathbb{N}_{0}}$ and
$\{l_{\mathrm{M}}(i)\}_{i\in\mathbb{N}_{0}}$ may be dependent. This
is clearly the case when the attacker has information of the random
packet losses in the channel. Furthermore, as we discussed in the
Introduction, the attacker may decide to attack based on the content
of packets. In such cases $l_{\mathrm{M}}(\cdot)$ would depend on
state and control input, which in turn depend on $l_{\mathrm{R}}(\cdot)$.
Proposition~\ref{PropositionDependentCombinedCase} below deals with
such cases. 

\begin{proposition} \label{PropositionDependentCombinedCase} Consider
the packet exchange failure indicator process $\{l(i)\in\{0,1\}\}_{i\in\mathbb{N}_{0}}$.
Assume 
\begin{align}
p_{1}+\rho_{\mathrm{M}} & <1,\label{eq:prop25cond}
\end{align}
where $p_{1}\in(0,1)$ is a scalar that satisfies (\ref{eq:p1condition-1}).
Then (\ref{eq:lcond1}) holds for all $\rho\in(p_{1}+\rho_{\mathrm{M}},1)$. 

\end{proposition} 

\begin{IEEEproof}It follows from (\ref{eq:ldefinitionforcombinedcase})
that 
\begin{align*}
L(k) & \leq\sum_{i=0}^{k-1}l_{\mathrm{R}}(i)+\sum_{i=0}^{k-1}l_{\mathrm{M}}(i),\quad k\in\mathbb{N}.
\end{align*}
Now, using arguments similar to the ones used for obtaining (\ref{eq:keyprobabilityinequality})
in the proof of Proposition~\ref{PropositionCombinedCase}, we have
\begin{align}
 & \mathbb{P}[L(k)>\rho k]\leq\mathbb{P}[\sum_{i=0}^{k-1}l_{\mathrm{R}}(i)+\sum_{i=0}^{k-1}l_{\mathrm{M}}(i)>\rho k]\nonumber \\
 & \,\,\leq\mathbb{P}[\sum_{i=0}^{k-1}l_{\mathrm{R}}(i)>\rho_{1}k]+\mathbb{P}[\sum_{i=0}^{k-1}l_{\mathrm{M}}(i)>\rho_{2}k],\,\,k\in\mathbb{N},\label{eq:separatedprobabilityterms}
\end{align}
 and consequently 
\begin{align}
 & \sum_{k=1}^{\infty}\mathbb{P}[L(k)>\rho k]\nonumber \\
 & \,\,\leq\sum_{k=1}^{\infty}\mathbb{P}[\sum_{i=0}^{k-1}l_{\mathrm{R}}(i)>\rho_{1}k]+\sum_{k=1}^{\infty}\mathbb{P}[\sum_{i=0}^{k-1}l_{\mathrm{M}}(i)>\rho_{2}k],\label{eq:sum-inequality-for-r-and-m}
\end{align}
where $\rho_{1}\triangleq p_{1}+\frac{\epsilon}{2}$, $\rho_{2}\triangleq\rho_{\mathrm{M}}+\frac{\epsilon}{2}$,
and $\epsilon\triangleq\rho-p_{1}-\rho_{\mathrm{M}}$. 

Observe that $\rho_{1}=p_{1}+\frac{\rho-p_{1}-\rho_{\mathrm{M}}}{2}=\frac{\rho+p_{1}-\rho_{\mathrm{M}}}{2}.$
Since $\frac{\rho+p_{1}-\rho_{\mathrm{M}}}{2}<1$ and $\epsilon>0$,
we have $\rho_{1}\in(p_{1},1)$. By using Lemma~\ref{Lemma-Random-Loss}
with $\rho_{\mathrm{R}}=\rho_{1}$, we obtain 
\begin{align}
\sum_{k=1}^{\infty}\mathbb{P}[L_{1}(k) & >\rho_{1}k]<\infty.\label{eq:first-sum-r}
\end{align}
Furthermore, note that $\rho_{2}=\rho_{\mathrm{M}}+\frac{\rho-p_{1}-\rho_{\mathrm{M}}}{2}=\frac{\rho+\rho_{\mathrm{M}}-p_{1}}{2}.$
Also, by $\frac{\rho+\rho_{\mathrm{M}}-p_{1}}{2}<1$ and $\epsilon>0$,
we have $\rho_{2}\in(\rho_{\mathrm{M}},1)$. Since $\rho_{2}>\rho_{\mathrm{M}}$,
by the characterization of $\{l_{\mathrm{M}}(i)\in\{0,1\}\}_{i\in\mathbb{N}_{0}}$,
\begin{align}
\sum_{k=1}^{\infty}\mathbb{P}[\sum_{i=0}^{k-1}l_{\mathrm{M}}(i)>\rho_{2}k] & \leq\sum_{k=1}^{\infty}\mathbb{P}[\sum_{i=0}^{k-1}l_{\mathrm{M}}(i)>\rho_{\mathrm{M}}k]<\infty.\label{eq:second-sum-m}
\end{align}
The result then follows from (\ref{eq:sum-inequality-for-r-and-m})--(\ref{eq:second-sum-m}).
\end{IEEEproof} 

In comparison with Proposition~\ref{PropositionCombinedCase}, the
result above provides a more restricted range of values for $\rho$
that satisfies Assumption~\ref{MainAssumption}. This is because
in Proposition~\ref{PropositionDependentCombinedCase} we find $\rho$
for the worst case scenario where the attacker may be knowledgeable
about all random packet losses in the network and may have access
to the information of the transmitted state and control input vectors.
An example scenario is where the attacker avoids placing malicious
attacks when there is already a random packet loss, increasing the
total number of packet exchange failures, which is clearly to the
disadvantage of the controller to maintain closed-loop stability. 

We note that the condition (\ref{eq:prop25cond}) guarantees that
the range $\rho\in(p_{1}+\rho_{\mathrm{M}},1)$ identified in Proposition~\ref{PropositionDependentCombinedCase}
is well defined. If $p_{1}+\rho_{M}\geq1$, then Assumption~\ref{MainAssumption}
holds with $\rho=1$. We also note that Proposition~\ref{PropositionDependentCombinedCase}
may introduce some conservativeness when it is applied to other scenarios
where malicious attacks and random packet losses are dependent, but
not as in the worst case scenario mentioned above. In such cases additional
information about the malicious attacks and random packet losses may
be employed to show that Assumption~\ref{MainAssumption} holds with
$\rho<1$ even if $p_{1}+\rho_{\mathrm{M}}\geq1$. 

\begin{remark} There may be situations where the attacker has limited
knowledge. For instance, the attacker may have access only to certain
entries of the state and control input vectors. This situation arises
in a multi-hop network with multiple paths (see, e.g., \cite{d2013fault,smarra2015});
different parts of the state and control input vectors may be sent
over different paths on the network and the attacker may have access
to the data only on some of those paths. In this case the attacker
would need an estimation mechanism to have information about the state/control
input vectors. Note that the operator may also utilize encryption
methods to prevent the attacker gain any information about the system
behavior. In the situations where the attacker is not knowledgeable
about the random packet losses and has no information of state and
control input vectors, Proposition~\ref{PropositionCombinedCase}
can be used. \end{remark}

\section{Event-Triggered Control Design \label{sec:Event-Triggered-Control-Design}}

In this section we investigate event-triggered control of (\ref{eq:system})
over an unreliable and potentially attacked network characterized
through Assumption~\ref{MainAssumption}. 

As a first step, we introduce the event-triggering scheme for communication
between the plant and the controller. This scheme will determine the
time instants $\tau_{i}\in\mathbb{N}_{0}$, $i\in\mathbb{N}_{0}$,
at which packet exchanges are attempted. For this purpose, we utilize
the quadratic Lyapunov-like function $V\colon\mathbb{R}^{n}\to[0,\infty)$
given by $V(x)\triangleq x^{\mathrm{T}}Px$, where $P>0$. Letting
$\tau_{0}=0$, we describe $\tau_{i}$, $i\in\mathbb{N}$, by 
\begin{align}
\tau_{i+1} & \triangleq\min\Big\{ t\in\{\tau_{i}+1,\tau_{i}+2,\ldots\}\colon t\geq\tau_{i}+\theta\nonumber \\
 & \quad\quad\quad\mathrm{or}\,\,\,V(Ax(t)+Bu(\tau_{i}))>\beta V(x(\tau_{i}))\Big\},\label{eq:attemptedpacketexchangetimes}
\end{align}
 where $\beta\in(0,1)$, $\theta\in\mathbb{N}$. 

The triggering condition (\ref{eq:attemptedpacketexchangetimes})
involves two parts. The part $V(Ax(t)+Bu(\tau_{i}))>\beta V(x(\tau_{i}))$
ensures that after a successful packet exchange attempt at $\tau_{i}$,
the value of $V(\cdot)$ stays below the level $\beta V(x(\tau_{i}))$
until the next packet exchange attempt. Furthermore, the triggering
condition $t\geq\tau_{i}+\theta$ ensures that two consecutive packet
exchange attempt instants are at most $\theta\in\mathbb{N}$ steps
apart, that is, $\tau_{i+1}-\tau_{i}\leq\theta$, $i\in\mathbb{N}_{0}$.
Although the specific value of $\theta$ does not affect the results
developed below, the boundedness of packet exchange attempt intervals
guarantees that $\tau_{i}$ (and hence $V(x(\tau_{i}))$) is well-defined
for each $i\in\mathbb{N}$. In practice, the value of $\theta$ can
be selected considering how frequent the plant state is desired to
be monitored by the controller side. 

\begin{figure}
\centering  \includegraphics[width=0.95\columnwidth]{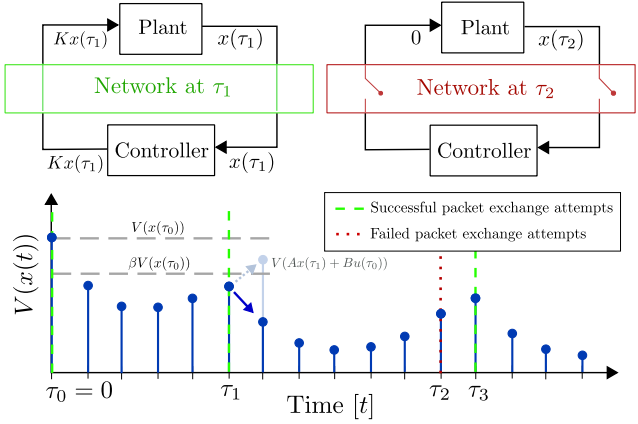} 

\caption{{[}Top{]} Networked control system with successful (left) and failed
(right) packet transmissions. \newline {[}Bottom{]} Response of the
Lyapunov-like function.}
 \vskip -10pt \label{operation}
\end{figure}

The operation of the event-triggered networked control system is illustrated
in Fig.~\ref{operation}. The triggering condition (\ref{eq:attemptedpacketexchangetimes})
is checked at the plant side at each step $t\in\mathbb{N}_{0}$. At
times $t=\tau_{i}$, $i\in\mathbb{N}$, the triggering condition is
satisfied and packet exchanges are attempted. In this example, a packet
exchange is attempted at time $t=\tau_{1}$, since $V(Ax(t)+Bu(\tau_{0}))>\beta V(x(\tau_{0}))$.
At this time instant, the plant and the controller successfully exchange
state and control input packets over the network, and as a result,
control input on the plant side is updated to $Kx(\tau_{1})$. Note
that packet exchange  attempts are not always successful, and may
fail due to loss of packets in the network. In the figure, the packet
exchange attempt at time $\tau_{2}$ fails. In this case, it follows
from (\ref{eq:control-input}) with $l(2)=1$ that the control input
at the plant side is set to $0$ at time $\tau_{2}$, which results
in an unstable behavior. A packet exchange is attempted again at the
very next time step $\tau_{3}$, since the triggering condition is
also satisfied at that time instant.

\subsection{Stability Analysis}

Next, we investigate stability of the closed-loop event-triggered
networked control system (\ref{eq:system}), (\ref{eq:control-input}),
(\ref{eq:attemptedpacketexchangetimes}), which is a stochastic dynamical
system due to the probabilistic characterization of packet losses.
Below we define almost sure asymptotic stability for stochastic dynamical
systems. 

\begin{definition}The zero solution $x(t)\equiv0$ of the stochastic
system (\ref{eq:system}), (\ref{eq:control-input}), and (\ref{eq:attemptedpacketexchangetimes})
is \emph{almost surely stable} if, for all $\epsilon>0$ and $\bar{p}>0$,
there exists $\delta=\delta(\epsilon,\bar{p})>0$ such that if $\|x(0)\|<\delta$,
then 
\begin{align}
\mathbb{P}[\sup_{t\in\mathbb{N}_{0}}\|x(t)\|>\epsilon] & <\bar{p}.\label{eq:almost-sure-stability}
\end{align}
 Moreover, the zero solution $x(t)\equiv0$ is \emph{asymptotically
stable almost surely} if it is almost surely stable and 
\begin{align}
\mathbb{P}[\lim_{t\to\infty}\|x(t)\|=0] & =1.\label{eq:definition-convergence}
\end{align}

\end{definition} 

In our stability analysis for the networked control system (\ref{eq:system}),
(\ref{eq:control-input}), we utilize an upper bound for the long
run average of the total number of failed packet exchanges. The following
result is a direct consequence of the Borel-Cantelli lemma (see \cite{klenke2008})
and shows that under Assumption~\ref{MainAssumption}, the long run
average of the total number of failed packet exchanges is upper bounded
by $\rho$ characterized in (\ref{eq:lcond1}). 

\begin{lemma} \label{key-lemma1} If there exists a scalar $\rho\in[0,1]$
such that (\ref{eq:lcond1}) holds, then 
\begin{align}
\limsup_{k\to\infty}\frac{L(k)}{k} & \leq\rho,\label{eq:long-run-average-bound}
\end{align}
almost surely. \end{lemma} 

In Propositions~\ref{PropositionCombinedCase} and \ref{PropositionDependentCombinedCase},
we obtained a range of values for $\rho$ that satisfy (\ref{eq:lcond1}).
In those results the range was given as an open interval. In the following
result we show that when Assumption~\ref{MainAssumption} holds for
a range of values, then (\ref{eq:long-run-average-bound}) also holds
with $\rho$ given as the infimum of the range. 

\begin{lemma} \label{Lemma-underbar-rho} Suppose (\ref{eq:lcond1})
is satisfied for all $\rho\in(\underline{\rho},1)$ where $\underline{\rho}\in[0,1)$.
Then (\ref{eq:long-run-average-bound}) holds with $\rho=\underline{\rho}$,
almost surely. \end{lemma}

\begin{IEEEproof}The proof resembles the sufficiency part of the
proof of Proposition 5.6 in \cite{karrprobabilitybook}. First, by
Lemma~\ref{key-lemma1}, 
\begin{align}
\mathbb{P}[\limsup_{k\to\infty}\frac{L(k)}{k}-\underline{\rho}>\epsilon] & =0,\label{eq:lemma-epsilon-version}
\end{align}
for any $\epsilon>0$. Now, it follows from (\ref{eq:lemma-epsilon-version})
that 
\begin{align*}
 & \mathbb{P}[\limsup_{k\to\infty}\frac{L(k)}{k}-\underline{\rho}>0]=\mathbb{P}[\cup_{j=1}^{\infty}\{\limsup_{k\to\infty}\frac{L(k)}{k}-\underline{\rho}>\frac{1}{j}\}]\\
 & \quad\quad\leq\sum_{j=1}^{\infty}\mathbb{P}[\limsup_{k\to\infty}\frac{L(k)}{k}-\underline{\rho}>\frac{1}{j}]=0,
\end{align*}
which implies that $\mathbb{P}[\limsup_{k\to\infty}\frac{L(k)}{k}\leq\underline{\rho}]=1$.
\end{IEEEproof} 

\begin{remark} Note that the term $\limsup_{k\to\infty}\frac{L(k)}{k}$
in (\ref{eq:long-run-average-bound}) corresponds to the ``discrete
event rate'' used in \cite{hassibi1999control,zhang2001stability}
for deterministic systems, when $\lim_{k\to\infty}\frac{L(k)}{k}$
exists. In this paper, Assumption~\ref{MainAssumption} allows the
binary-valued process $\{l(i)\in\{0,1\}\}_{i\in\mathbb{N}_{0}}$ to
be a non-ergodic stochastic process, for which $\lim_{k\to\infty}\frac{L(k)}{k}$
may not be equal for all sample paths. For instance, let $l(i)\triangleq l_{\mathrm{M}}(i),i\in\mathbb{N}_{0}$,
 and 
\begin{align*}
l_{\mathrm{M}}(i) & \triangleq\begin{cases}
1,\quad & i\in\{\alpha,2\alpha,3\alpha\ldots\},\\
0,\quad & \mathrm{otherwise},
\end{cases}
\end{align*}
 where $\alpha:\Omega\to\{2,4\}$ is a random variable with $\mathbb{P}[\alpha=2]=\mathbb{P}[\alpha=4]=\frac{1}{2}$.
In this setting, the attacker decides the period of attacks based
on a random variable $\alpha:\Omega\to\{2,4\}$. Depending on the
value of $\alpha$, malicious packet losses occur either at every
$2$ packet exchange attempts or at every $4$ packet exchange attempts.
Thus, the discrete event rate would be a random variable that depends
on the value of $\alpha$. On the other hand, regardless of the value
of $\alpha$, (\ref{eq:jammingcondition}) is satisfied with $\tau=2$,
and hence Lemmas~\ref{JammingLemma} and \ref{Lemma-underbar-rho}
imply that $\limsup_{k\to\infty}\frac{L(k)}{k}\leq\frac{1}{2},$ almost
surely. Note that here $\frac{1}{2}$ represents the worst-case upper
bound for the long run average of the total number of failed packet
exchanges. \end{remark}

We are now ready to state the main result of this paper. It provides
a sufficient condition for almost sure asymptotic stability of the
networked control system (\ref{eq:system}), (\ref{eq:control-input})
with packet exchange failure indicator $\{l(i)\in\{0,1\}\}_{i\in\mathbb{N}_{0}}$
satisfying (\ref{eq:long-run-average-bound}). 

\begin{theorem} \label{TheoremMain} Consider the linear dynamical
system (\ref{eq:system}). Suppose that the process $\{l(i)\in\{0,1\}\}_{i\in\mathbb{N}_{0}}$
characterizing packet exchange failures\begin{footnote}{We set (\ref{eq:long-run-average-bound})
as a condition for packet exchange failures as it allows more generality
in comparison to Assumption~\ref{MainAssumption}. Note that by Lemma~\ref{key-lemma1},
Assumption~\ref{MainAssumption} implies (\ref{eq:long-run-average-bound}).
Furthermore, Lemma~\ref{Lemma-underbar-rho} shows that (\ref{eq:long-run-average-bound})
also holds when $\rho$ is given as the infimum of an open interval
where all values satisfy Assumption~\ref{MainAssumption}.}{}\end{footnote}
in the network satisfies (\ref{eq:long-run-average-bound}) with scalar
$\rho\in[0,1]$. If there exist a matrix $K\in\mathbb{R}^{m\times n}$,
a positive-definite matrix $P\in\mathbb{R}^{n\times n}$, and scalars
$\beta\in(0,1),$ $\varphi\in[1,\infty)$ such that 
\begin{align}
 & \left(A+BK\right)^{\mathrm{T}}P\left(A+BK\right)-\beta P\leq0,\label{eq:betacond}\\
 & A^{\mathrm{T}}PA-\varphi P\leq0,\label{eq:varphicond}\\
 & (1-\rho)\ln\beta+\rho\ln\varphi<0,\label{eq:betaandvarphicond}
\end{align}
then the event-triggered control law (\ref{eq:control-input}), (\ref{eq:attemptedpacketexchangetimes})
guarantees almost sure asymptotic stability of the zero solution $x(t)\equiv0$
of the closed-loop system dynamics. 

\end{theorem}

\begin{IEEEproof}The proof is composed of three steps. In the initial
step, we obtain an inequality concerning the evolution of the Lyapunov-like
function $V(x)\triangleq x^{\mathrm{T}}Px$, $x\in\mathbb{R}^{n}$.
Then, we will establish almost sure stability, and then finally we
show almost sure asymptotic stability of the closed-loop system. 

First, we use (\ref{eq:system}) and (\ref{eq:control-input}) together
with $V(\cdot)$ to obtain 
\begin{align}
V(x(\tau_{i}+1)) & =x^{\mathrm{T}}(\tau_{i})\left(A+\left(1-l(i)\right)BK\right)^{\mathrm{T}}P\nonumber \\
 & \quad\,\cdot\left(A+\left(1-l(i)\right)BK\right)x(\tau_{i}),\,i\in\mathbb{N}_{0}.\label{eq:vattauplush}
\end{align}
 Now, for the case $l(i)=0$, (\ref{eq:betacond}) and (\ref{eq:vattauplush})
imply 
\begin{align}
V(x(\tau_{i}+1)) & =x^{\mathrm{T}}(\tau_{i})\left(A+BK\right)^{\mathrm{T}}P\left(A+BK\right)x(\tau_{i})\nonumber \\
 & \leq\beta x^{\mathrm{T}}(\tau_{i})Px(\tau_{i}).\label{eq:betaineq}
\end{align}
 Since $\tau_{i+1}\geq\tau_{i}+1$, it follows from (\ref{eq:attemptedpacketexchangetimes})
and (\ref{eq:betaineq}) that 
\begin{align}
V(x(t)) & \leq\beta x^{\mathrm{T}}(\tau_{i})Px(\tau_{i})\nonumber \\
 & =\beta V(x(\tau_{i})),\quad t\in\{\tau_{i}+1,\ldots,\tau_{i+1}\}.\label{eq:betaresult}
\end{align}

On the other hand, for the case $l(i)=1$, we have from (\ref{eq:varphicond})
and (\ref{eq:vattauplush}) that 
\begin{align}
V(x(\tau_{i}+1)) & =x^{\mathrm{T}}(\tau_{i})A^{\mathrm{T}}PAx(\tau_{i})\leq\varphi x^{\mathrm{T}}(\tau_{i})Px(\tau_{i}).\label{eq:varphiineq}
\end{align}
Now if $\tau_{i+1}=\tau_{i}+1$, we have $V(x(\tau_{i+1}))\leq\varphi V(x(\tau_{i}))$
due to (\ref{eq:varphiineq}). Otherwise, that is, if $\tau_{i+1}>\tau_{i}+1$,
it means that $V(x(t))\leq\beta V(x(\tau_{i}))$ for $t\in\{\tau_{i}+2,\ldots,\tau_{i+1}\}$.
Therefore, since $\beta\leq\varphi$, 
\begin{align}
V(x(t)) & \leq\varphi V(x(\tau_{i})),\quad t\in\{\tau_{i}+1,\ldots,\tau_{i+1}\}.\label{eq:varphiresult}
\end{align}
Using (\ref{eq:betaresult}) and (\ref{eq:varphiresult}) we obtain
\begin{align}
V(x(\tau_{i+1})) & \leq(1-l(i))\beta V(x(\tau_{i}))+l(i)\varphi V(x(\tau_{i})),\label{eq:vineq}
\end{align}
 for $i\in\mathbb{N}_{0}$. Note that the inequality given in (\ref{eq:vineq})
provides an upper bound on $V(\cdot)$. 

Now, let $\eta(k)\triangleq\prod_{i=0}^{k-1}\left[(1-l(i))\beta+l(i)\varphi\right]$.
Then, by (\ref{eq:vineq}), 
\begin{align}
V(x(\tau_{k})) & \leq\eta(k)V(x(0)),\quad k\in\mathbb{N}.\label{eq:vetaineq}
\end{align}
 Furthermore, since $\ln\left[(1-q)\beta+q\varphi\right]=(1-q)\ln\beta+q\ln\varphi$
for $q\in\{0,1\}$, we have 
\begin{align*}
\ln\eta(k) & =\sum_{i=0}^{k-1}\ln\left[(1-l(i))\beta+l(i)\varphi\right]\\
 & =\sum_{i=0}^{k-1}(1-l(i))\ln\beta+\sum_{i=0}^{k-1}l(i)\ln\varphi\\
 & =(k-L(k))\ln\beta+L(k)\ln\varphi,
\end{align*}
 where $L(k)=\sum_{i=0}^{k-1}l(i)$ by (\ref{eq:bigldefinition}).
Now by $\beta\in(0,1)$, and $\varphi\in[1,\infty)$, it follows from
(\ref{eq:long-run-average-bound}) and (\ref{eq:betaandvarphicond})
that 
\begin{align*}
\limsup_{k\to\infty}\frac{\ln\eta(k)}{k} & =\limsup_{k\to\infty}\frac{1}{k}\left[(k-L(k))\ln\beta+L(k)\ln\varphi\right]\\
 & \leq(1-\rho)\ln\beta+\rho\ln\varphi<0,
\end{align*}
 almost surely. As a consequence, $\lim_{k\to\infty}\ln\eta(k)=-\infty$,
and hence, $\lim_{k\to\infty}\eta(k)=0$, almost surely. Thus, for
any $\epsilon>0$, $\lim_{j\to\infty}\mathbb{P}[\sup_{k\geq j}\eta(k)>\epsilon^{2}]=0$.
Therefore, for any $\epsilon>0$ and $\bar{p}>0$, there exists a
positive integer $N(\epsilon,\bar{p})$ such that 
\begin{align}
\mathbb{P}[\sup_{k\geq j}\eta(k) & >\epsilon^{2}]<\bar{p},\quad j\geq N(\epsilon,\bar{p}).\label{eq:supetainequalit}
\end{align}

In what follows, we employ (\ref{eq:vetaineq}) and (\ref{eq:supetainequalit})
to show almost sure stability of the closed-loop system. Note that
(\ref{eq:betaresult}), (\ref{eq:varphiresult}), and $\varphi\geq1>\beta$
imply that $V(x(t+1))\leq\varphi V(x(t)),\,t\in\{\tau_{i},\ldots,\tau_{i+1}-1\},$
$i\in\mathbb{N}_{0}$. Since $\|x\|^{2}\leq\frac{1}{\lambda_{\min}(P)}V(x)$
and $V(x)\leq\lambda_{\max}(P)\|x\|^{2}$, $x\in\mathbb{R}^{n}$,
we have 
\begin{align}
\|x(t)\|^{2} & \leq\varphi\nu\|x(\tau_{i})\|^{2},\,t\in\{\tau_{i},\ldots,\tau_{i+1}-1\}\label{eq:xtnuineq}
\end{align}
for $i\in\mathbb{N}_{0}$, where $\nu\triangleq\frac{\lambda_{\max}(P)}{\lambda_{\min}(P)}$. 

Now, let $\mathcal{T}_{k}\triangleq\{\tau_{k},\ldots,\tau_{k+1}-1\}$,
$k\in\mathbb{N}_{0}$. Then by using (\ref{eq:vetaineq}) and (\ref{eq:xtnuineq}),
we obtain $\eta(k)\geq\frac{V(x(\tau_{k}))}{V(x(0))}\geq\frac{\lambda_{\min}(P)}{\lambda_{\max}(P)}\frac{\|x(\tau_{k})\|^{2}}{\|x(0)\|^{2}}\geq\frac{1}{\nu^{2}\varphi}\frac{\|x(t)\|^{2}}{\|x(0)\|^{2}}$
for all $t\in\mathcal{T}_{k}$, $k\in\mathbb{N}$. Hence, $\eta(k)\geq\frac{1}{\nu^{2}\varphi}\frac{\max_{t\in\mathcal{T}_{k}}\|x(t)\|^{2}}{\|x(0)\|^{2}}$,
$k\in\mathbb{N}$. By (\ref{eq:supetainequalit}), it follows that
for all $\epsilon>0$ and $\bar{p}>0$, 
\begin{align*}
 & \mathbb{P}[\sup_{k\geq j}\max_{t\in\mathcal{T}_{k}}\|x(t)\|>\epsilon\nu\sqrt{\varphi}\|x(0)\|]\\
 & \quad=\mathbb{P}[\sup_{k\geq j}\max_{t\in\mathcal{T}_{k}}\|x(t)\|^{2}>\epsilon^{2}\nu^{2}\varphi\|x(0)\|^{2}]\\
 & \quad=\mathbb{P}[\sup_{k\geq j}\frac{1}{\nu^{2}\varphi}\frac{\max_{t\in\mathcal{T}_{k}}\|x(t)\|^{2}}{\|x(0)\|^{2}}>\epsilon^{2}]\\
 & \quad\leq\mathbb{P}[\sup_{k\geq j}\eta(k)>\epsilon^{2}]<\bar{p},\quad j\geq N(\epsilon,\bar{p}).
\end{align*}
We now define $\delta_{1}\triangleq\frac{1}{\nu\sqrt{\varphi}}$.
Note that if $\|x(0)\|\leq\delta_{1}$, then (since $\nu\sqrt{\varphi}\|x(0)\|\leq1$)
for all $j\geq N(\epsilon,\bar{p})$, we have 
\begin{align}
 & \mathbb{P}[\sup_{k\geq j}\max_{t\in\mathcal{T}_{k}}\|x(t)\|>\epsilon]\nonumber \\
 & \quad\leq\mathbb{P}[\sup_{k\geq j}\max_{t\in\mathcal{T}_{k}}\|x(t)\|>\epsilon\nu\sqrt{\varphi}\|x(0)\|]<\bar{p}.\label{eq:epsilon-result-part1}
\end{align}
 On the other hand, since $\varphi\geq1>\beta$, it follows from (\ref{eq:vineq})
that $V(x(\tau_{k}))\leq\varphi^{k}V(x(0))\leq\varphi^{N(\epsilon,\bar{p})-1}V(x(0))$
for all $k\in\{0,1,\ldots,N(\epsilon,\bar{p})-1\}$. Therefore, $\|x(\tau_{k})\|^{2}\leq\varphi^{N(\epsilon,\bar{p})-1}\frac{\lambda_{\max}(P)}{\lambda_{\min}(P)}\|x(0)\|^{2}=\varphi^{N(\epsilon,\bar{p})-1}\nu\|x(0)\|^{2}$.
Furthermore, as a result of (\ref{eq:xtnuineq}), 
\begin{align*}
\max_{t\in\mathcal{T}_{k}}\|x(t)\|^{2} & \leq\varphi\nu\|x(\tau_{k})\|^{2}\leq\nu^{2}\varphi^{N(\epsilon,\bar{p})}\|x(0)\|^{2},
\end{align*}
 and hence, $\max_{t\in\mathcal{T}_{k}}\|x(t)\|\leq\nu\sqrt{\varphi^{N(\epsilon,\bar{p})}}\|x(0)\|$
for all $k\in\{0,1,\ldots,N(\epsilon,\bar{p})-1\}$. Let $\delta_{2}\triangleq\epsilon\nu^{-1}\sqrt{\varphi^{-N(\epsilon,\bar{p})}}$.
Now, if $\|x(0)\|\leq\delta_{2}$, then $\max_{t\in\mathcal{T}_{k}}\|x(t)\|\leq\epsilon$,
$k\in\{0,1,\ldots,N(\epsilon,\bar{p})-1\}$, which implies 
\begin{eqnarray}
\mathbb{P}[\max_{k\in\{0,1,\ldots,N(\epsilon,\bar{p})\}}\max_{t\in\mathcal{T}_{k}}\|x(t)\|>\epsilon] & = & 0.\label{eq:epsilon-result-part2}
\end{eqnarray}
It follows from (\ref{eq:epsilon-result-part1}) and (\ref{eq:epsilon-result-part2})
that for all $\epsilon>0$, $\bar{p}>0$, 
\begin{align*}
 & \mathbb{P}[\sup_{t\in\mathbb{N}_{0}}\|x(t)\|>\epsilon]=\mathbb{P}[\sup_{k\in\mathbb{N}_{0}}\max_{t\in\mathcal{T}_{k}}\|x(t)\|>\epsilon]\\
 & \quad=\mathbb{P}[\{\max_{k\in\{0,1,\ldots,N(\epsilon,\bar{p})-1\}}\max_{t\in\mathcal{T}_{k}}\|x(t)\|>\epsilon\}\\
 & \quad\quad\quad\cup\,\{\sup_{k\geq N(\epsilon,\bar{p})}\max_{t\in\mathcal{T}_{k}}\|x(t)\|>\epsilon\}]\\
 & \quad\leq\mathbb{P}[\max_{k\in\{0,1,\ldots,N(\epsilon,\bar{p})-1\}}\max_{t\in\mathcal{T}_{k}}\|x(t)\|>\epsilon]\\
 & \quad\quad+\mathbb{P}[\sup_{k\geq N(\epsilon,\bar{p})}\max_{t\in\mathcal{T}_{k}}\|x(t)\|>\epsilon]<\bar{p},
\end{align*}
 whenever $\|x(0)\|<\delta\triangleq\min(\delta_{1},\delta_{2})$,
which implies almost sure stability. 

Finally, in order to establish almost sure \emph{asymptotic} stability
of the zero solution, it remains to show (\ref{eq:definition-convergence}).
To this end, observe that $\mathbb{P}[\lim_{k\to\infty}\eta(\tau_{k})=0]=1$.
It follows from (\ref{eq:vetaineq}) that $\mathbb{P}[\lim_{k\to\infty}V(x(\tau_{k}))=0]=1$,
which implies (\ref{eq:definition-convergence}). Hence the zero solution
of the closed-loop system (\ref{eq:system}), (\ref{eq:control-input}),
(\ref{eq:attemptedpacketexchangetimes}) is asymptotically stable
almost surely.\end{IEEEproof} 

Theorem~\ref{TheoremMain} provides a sufficient condition under
which the event-triggered control law (\ref{eq:control-input}), (\ref{eq:attemptedpacketexchangetimes})
guarantees almost sure asymptotic stability of the system (\ref{eq:system})
for the case of packet losses satisfying Assumption~\ref{MainAssumption}.
Note that the scalars $\beta\in(0,1)$ and $\varphi\in[1,\infty)$
in conditions (\ref{eq:betacond}) and (\ref{eq:varphicond}) characterize
upper bounds on the growth of the Lyapunov-like function, and they
are also related to closed-loop and open-loop bounds utilized in \cite{kellett2005stability,quevedo2014stochastic}.
Specifically, when a packet exchange attempt between the plant and
the controller is successful at time $\tau_{i}$, the condition (\ref{eq:betacond})
together with (\ref{eq:attemptedpacketexchangetimes}) guarantees
that $V(x(\tau_{i+1}))\leq\beta V(x(\tau_{i}))$. On the other hand,
if a packet exchange is unsuccessful at time $\tau_{i}$, it follows
from (\ref{eq:attemptedpacketexchangetimes}) and (\ref{eq:varphicond})
that $V(x(\tau_{i+1}))\leq\varphi V(x(\tau_{i}))$. If successful
packet exchanges are sufficiently frequent such that (\ref{eq:betaandvarphicond})
is satisfied, then the closed-loop stability is guaranteed. 

We remark that the analysis for the closed-loop system stability in
the proof above is technically involved partly due to the general
characterization in Assumption~\ref{MainAssumption}, which captures
not only random packet losses but attacks as well. If we consider
only random packet losses, we may employ methods from discrete-time
Markov jump systems theory \cite{costa2004discrete} for obtaining
conditions of stability. Furthermore, in the case $\{l(i)\in\{0,1\}\}_{i\in\mathbb{N}_{0}}$
is an ergodic process, the results presented in \cite{kellett2005stability}
can be directly employed to show stability. 

On the other hand packet losses due to attacks (Section~\ref{RemarkJammingAttacks})
cannot be described using Markov processes and they may not be ergodic.
Stability of a system under denial-of-service attacks is explored
in \cite{depersis2014}, where the analysis relies on a deterministic
approach for obtaining an exponentially decreasing upper bound for
the norm of the state. In contrast, in our analysis, we use probabilistic
approaches similar to \cite{kellett2005stability,Lemmon:2011:ASS:1967701.1967744}
to show almost sure asymptotic stability. Specifically, we use tools
from probability theory to find a stochastic upper bound for a Lyapunov-like
function and show that this bound tends to zero even though it may
increase at certain times. This approach is related to obtaining an
upper bound on the \emph{top Lyapunov exponent} (see \cite{fang1995stability,ezzine1997almost,bolzern2004almost})
of a stochastic system. 

Theorem~\ref{TheoremMain} provides conditions that guarantee both
(\ref{eq:almost-sure-stability}) and (\ref{eq:definition-convergence})
implying almost sure asymptotic stability. In this stability definition,
(\ref{eq:definition-convergence}) is concerned with the convergence
of solutions to zero, while (\ref{eq:almost-sure-stability}) ensures
that states sufficiently close to the origin are likely to stay close
to the origin. However note that (\ref{eq:almost-sure-stability})
allows states to leave any given ball in a finite time with positive
(even if small) probability. For instance, if many consecutive packet
transmission attempts fail, the state magnitude may grow due to lack
of control action. We emphasize that Assumption~\ref{MainAssumption}
and hence (\ref{eq:long-run-average-bound}) ensure packet failures
to be statistically rare so that the state eventually converges to
the origin. 

\begin{remark}In addition to almost sure stability, there are other
stochastic stability and performance notions that are useful for the
analysis of networked control systems. In particular, moment stability
and moment-based performance notions have been utilized when random
packet losses are considered (see \cite{hespanha2007,schenato2007}
and the references therein). In comparison with those works, in our
problem setting, we must take into account also the effect of malicious
attacks. We remark that in contrast with random packet losses, precise
information of the probabilities of malicious attacks is not available.
Hence, it is difficult to characterize the evolution of the moments
of the state and establish moment stability. On the other hand, both
random packet losses and malicious attacks, as well as their combination
provide us information about the asymptotic ratio of packet exchange
failures, which can be employed in the analysis when we consider almost
sure asymptotic stability. \end{remark}

In the following corollary of Theorem~\ref{TheoremMain}, we discuss
the special case of random packet losses described with time-homogeneous
Markov chains. 

\begin{corollary} \label{CorollaryRandom} Consider the linear dynamical
system (\ref{eq:system}). Suppose that the process $\{l(i)\in\{0,1\}\}_{i\in\mathbb{N}_{0}}$
is an irreducible time-homogeneous Markov chain with constant transition
probabilities $p_{q,r}\in[0,1]$, $q,r\in\{0,1\}$. If there exist
a matrix $K\in\mathbb{R}^{m\times n}$, a positive-definite matrix
$P\in\mathbb{R}^{n\times n}$, and scalars $\beta\in(0,1),$ $\varphi\in[1,\infty)$
such that (\ref{eq:betacond}), (\ref{eq:varphicond}) and (\ref{eq:betaandvarphicond})
hold with $\rho\triangleq\frac{p_{0,1}}{p_{0,1}+p_{1,0}}$, then the
event-triggered control law (\ref{eq:control-input}), (\ref{eq:attemptedpacketexchangetimes})
guarantees almost sure asymptotic stability of the zero solution $x(t)\equiv0$
of the closed-loop system dynamics. \end{corollary}

\begin{IEEEproof} By the ergodic theorem for irreducible Markov chains
\cite{norris2009}, we have $\lim_{k\to\infty}\frac{L(k)}{k}=\rho$.
Now, since (\ref{eq:long-run-average-bound}) holds, the result follows
from Theorem~\ref{TheoremMain}. \end{IEEEproof}

When we consider transmission attempts at all times by setting $\theta=1$
in (\ref{eq:attemptedpacketexchangetimes}), Corollary~\ref{CorollaryRandom}
recovers a specialization of the result in \cite{kellett2005stability}
for linear systems. Furthermore, if we consider $\{l(i)\in\{0,1\}\}_{i\in\mathbb{N}_{0}}$
to be a Bernoulli process, then $\rho$ in Corollary~\ref{CorollaryRandom}
is given by $\rho=p$, where $p=p_{0,1}=p_{1,1}$ denotes the packet
loss probability. In this setting, the almost sure stability condition
in Corollary~\ref{CorollaryRandom} is tighter than the second-moment
stability condition in \cite{quevedo2014stochastic}. Specifically,
for this problem setting, the results in \cite{quevedo2014stochastic}
can be used to obtain the second-moment stability condition $(1-\rho)\beta+\rho\varphi<1$
or equivalently $\ln[(1-\rho)\beta+\rho\varphi]<0$. In comparison
to this condition, the stability condition (\ref{eq:betaandvarphicond})
in Corollary~\ref{CorollaryRandom} is tighter. This is because $(1-\rho)\ln\beta+\rho\ln\varphi<\ln[(1-\rho)\beta+\rho\varphi]$
by Jensen's inequality, since $\beta<\varphi$ and $\rho\notin\{0,1\}$.

\subsection{Feedback Gain Design for Event-Triggered Control \label{sub:Feedback-Gain-Design}}

In the following, we outline a numerical method for designing the
feedback gain $K\in\mathbb{R}^{m\times n}$, as well as the positive-definite
matrix $P\in\mathbb{R}^{n\times n}$ and the scalar $\beta\in(0,1)$
used in the event-triggered control law (\ref{eq:control-input}),
(\ref{eq:attemptedpacketexchangetimes}). 

\begin{corollary} \label{Corollary} Consider the linear dynamical
system (\ref{eq:system}). Suppose that the process $\{l(i)\in\{0,1\}\}_{i\in\mathbb{N}_{0}}$
characterizing packet exchange failures in the network satisfies (\ref{eq:long-run-average-bound})
with scalar $\rho\in[0,1]$. If there exist a matrix $M\in\mathbb{R}^{m\times n}$,
a positive-definite matrix $Q\in\mathbb{R}^{n\times n}$, and scalars
$\beta\in(0,1),$ $\varphi\in[1,\infty)$ such that (\ref{eq:betaandvarphicond}),
\begin{align}
\left[\begin{array}{cc}
\beta Q & \left(AQ+BM\right)^{\mathrm{T}}\\
AQ+BM & Q
\end{array}\right] & \geq0,\label{eq:corolcond1}\\
\left[\begin{array}{cc}
\varphi Q & (AQ){}^{\mathrm{T}}\\
AQ & Q
\end{array}\right] & \geq0,\label{eq:corolcond2}
\end{align}
hold, then the event-triggered control law (\ref{eq:control-input}),
(\ref{eq:attemptedpacketexchangetimes}) with $P\triangleq Q^{-1}$
and $K\triangleq MQ^{-1}$ guarantees almost sure asymptotic stability
of the zero solution $x(t)\equiv0$ of the closed-loop system dynamics. 

\end{corollary}

\begin{IEEEproof}Using Schur complements (see \cite{bernstein2009matrix}),
we transform (\ref{eq:corolcond1}) and (\ref{eq:corolcond2}), respectively,
into 
\begin{align}
\beta Q-\left(AQ+BM\right)^{\mathrm{T}}Q^{-1}\left(AQ+BM\right) & \geq0,\label{eq:oldcorolcond1}\\
\varphi Q-(AQ)^{\mathrm{T}}Q^{-1}AQ & \geq0.\label{eq:oldcorolcond2}
\end{align}
By multiplying both sides of (\ref{eq:oldcorolcond1}) and (\ref{eq:oldcorolcond2})
from left and right by $Q^{-1}$, we obtain (\ref{eq:betacond}) and
(\ref{eq:varphicond}) with $P=Q^{-1}$ and $K=MQ^{-1}$. Thus, the
result follows from Theorem~\ref{TheoremMain}. \end{IEEEproof}

\begin{figure}
\centering \includegraphics[width=0.65\columnwidth]{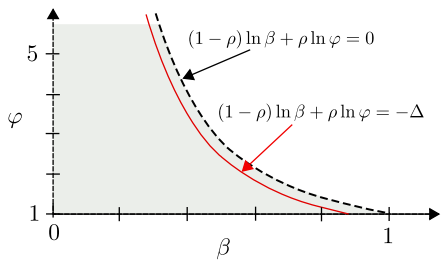}\vskip -7pt\caption{Region for $\beta\in(0,1)$ and $\varphi\in[1,\infty)$ that satisfy
(\ref{eq:betaandvarphicond}) for $\rho=0.4$ }
 \vskip -15pt \label{FigBetaandVarphi}
\end{figure}

We remark that the matrix inequalities (\ref{eq:corolcond1}) and
(\ref{eq:corolcond2}) are linear in $M\in\mathbb{R}^{m\times n}$
and $Q\in\mathbb{R}^{n\times n}$ for fixed $\beta\in(0,1)$ and $\varphi\in[1,\infty)$.
In our method we seek feasible solutions $M$ and $Q$ for linear
matrix inequalities (\ref{eq:corolcond1}) and (\ref{eq:corolcond2})
by iterating over a set of values for $\beta\in(0,1)$ and $\varphi\in[1,\infty)$
restricted by the condition (\ref{eq:betaandvarphicond}). It is however
noted that we do not need to search $\beta$ and $\varphi$ in the
entire range characterized by (\ref{eq:betaandvarphicond}). It turns
out to be sufficient to check for larger values of $\beta$ and $\varphi$
that are close to the boundary of the range identified by $(1-\rho)\ln\beta+\rho\ln\varphi=0$.
Specifically, we set $\Delta>0$ as a small positive real number,
and then we iterate over a set of values for $\beta$ in the range
$(0,e^{-\frac{\Delta}{1-\rho}}]$ to look for feasible solutions $M$
and $Q$ for the linear matrix inequalities (\ref{eq:corolcond1})
and (\ref{eq:corolcond2}) with $\varphi=e^{-\frac{(1-\rho)\ln\beta+\Delta}{\rho}}$.
In this approach, we use only $\beta\in(0,1)$, $\varphi\in[1,\infty)$
that are on the curve $(1-\rho)\ln\beta+\rho\ln\varphi=-\Delta$.
We illustrate this curve with the solid red line in Fig.~\ref{FigBetaandVarphi},
where the shaded region corresponds to $\beta$ and $\varphi$ that
satisfy (\ref{eq:betaandvarphicond}). Note that picking smaller values
for $\Delta>0$ moves the curve towards the boundary. Also, there
is no conservatism in \emph{not} considering $\beta$ and $\varphi$
such that $(1-\rho)\ln\beta+\rho\ln\varphi<-\Delta$. This is because
if there exist $M$ and $Q$ that satisfy (\ref{eq:corolcond1}) and
(\ref{eq:corolcond2}) for values $\beta=\tilde{\beta}$ and $\varphi=\tilde{\varphi}$,
then the same $M$ and $Q$ satisfy (\ref{eq:corolcond1}) and (\ref{eq:corolcond2})
also for larger values $\beta>\tilde{\beta}$ and $\varphi>\tilde{\varphi}$.

\section{Attacker's Perspective \label{sec:Attacker's-Perspective}}

In order to design \emph{cyber-secure} control systems, it is essential
to understand the risks in networked operation. In this regard, it
may be useful to consider the control problem from the perspective
of an attacker. An attacker knowledgeable about the networked control
system may generate an attack strategy that causes sufficiently frequent
packet losses which can result in instability of the closed-loop dynamics.
However, the attacker may want to keep the number of attacks as small
as possible. One reason in the case of jamming attacks is that monitoring
the channel and producing jamming signals consume energy \cite{xu2005feasibility}.
Moreover, the attacks should be kept minimal to make them less detectable
by the system operators. 

In this section, we address the question of finding conditions under
which the state diverges almost surely (i.e., $\mathbb{P}[\lim_{t\to\infty}\|x(t)\|=\infty]=1$).
For the discussions and results presented in this section, we consider
the case where the plant and the controller attempt to exchange packets
at all time instants, that is, $\tau_{i}=i$, $i\in\mathbb{N}_{0}$.
In the event-triggered scheme, this corresponds to the case with $\theta=1$
in (\ref{eq:attemptedpacketexchangetimes}). 

First, we obtain a \emph{lower-bound} for the long run average number
of packet exchange failures by utilizing a characterization that is
complementary to (\ref{eq:lcond1}) in Assumption~\ref{MainAssumption}. 

\begin{lemma} \label{key-lemma1-1} If there exists a scalar $\sigma\in[0,1]$
such that 
\begin{align}
\sum_{k=1}^{\infty}\mathbb{P}[L(k)<\sigma k] & <\infty,\label{eq:c1-1}
\end{align}
where $L(k)\triangleq\sum_{i=0}^{k-1}l(t)$, $k\in\mathbb{N}$, then
\begin{align}
\liminf_{k\to\infty}\frac{L(k)}{k} & \geq\sigma,\label{eq:long-run-average-bound-1}
\end{align}
almost surely. \end{lemma} 

\begin{IEEEproof} Using (\ref{eq:c1-1}), we obtain $\sum_{k=1}^{\infty}\mathbb{P}[\sum_{t=0}^{k-1}(1-l(t))>(1-\sigma)k]<\infty$,
and hence, by Borel-Cantelli lemma (see \cite{klenke2008}), 
\begin{align}
\limsup_{k\to\infty}\frac{1}{k}\sum_{t=0}^{k-1}(1-l(t)) & \leq1-\sigma,\label{eq:bc}
\end{align}
 almost surely. Noting that $\limsup_{k\to\infty}\frac{1}{k}\sum_{t=0}^{k-1}(1-l(t))=1-\liminf_{k\to\infty}\frac{1}{k}\sum_{t=0}^{k-1}l(t)$,
we obtain (\ref{eq:long-run-average-bound-1}) from (\ref{eq:bc}).
\end{IEEEproof}

The inequality (\ref{eq:c1-1}) can be considered as a complementary
characterization to (\ref{eq:lcond1}) in Assumption~\ref{MainAssumption}.
Observe that by Lemma~\ref{key-lemma1}, $\rho\in[0,1]$ in (\ref{eq:lcond1})
characterizes an \emph{upper-bound} on the long run average number
of packet exchange failures. In comparison, as implied by (\ref{eq:long-run-average-bound-1}),
the scalar $\sigma\in[0,1]$ in (\ref{eq:c1-1}) provides a \emph{lower-bound}
on the long run average number of packet exchange failures. 

Notice that a large $\sigma\in[0,1]$ in (\ref{eq:long-run-average-bound-1})
indicates that due to random losses and malicious attacks, packet
exchange failures happen statistically frequently. In such cases,
the overall dynamics may become unstable. As mentioned earlier, since
malicious attacks often consume energy, the attacker would want to
disrupt normal operation and cause unstable behavior with a fewer
number of attacks. In the case of jamming attacks, recent works considered
game-theoretic methods to investigate the optimal strategy of an attacker
when the jamming energy is a constraint in the problem \cite{li2015jamming}
and when it is part of the attacker's cost function \cite{alpcan2010network,liu2014stochastic}.
The results obtained there are not directly applicable here, as we
investigate sufficient attack rates that cause divergence of the state
rather than finding optimal attack strategies. 

Our next result indicates how frequently the attacker should cause
packet exchange failures to induce instability. 

\begin{theorem} \label{TheoremMain-Attackers-Perspective} Consider
the linear networked control system (\ref{eq:system}), (\ref{eq:control-input})
where packet exchanges between the plant and the controller are attempted
at all time instants. Suppose that the process $\{l(t)\in\{0,1\}\}_{t\in\mathbb{N}_{0}}$
characterizing packet exchange failures in the network satisfies (\ref{eq:long-run-average-bound-1})
with $\sigma\in[0,1]$. If there exist a positive-definite matrix
$\hat{P}\in\mathbb{R}^{n\times n}$ and scalars $\hat{\beta}\in(0,1),$
$\hat{\varphi}\in[1,\infty)$ such that 
\begin{align}
 & \left(A+BK\right)^{\mathrm{T}}\hat{P}\left(A+BK\right)-\hat{\beta}\hat{P}\geq0,\label{eq:betacond-1}\\
 & A^{\mathrm{T}}\hat{P}A-\hat{\varphi}\hat{P}\geq0,\label{eq:varphicond-1}\\
 & (1-\sigma)\ln\hat{\beta}+\sigma\ln\hat{\varphi}>0,\label{eq:betaandvarphicond-1}
\end{align}
then $\lim_{t\to\infty}\|x(t)\|=\infty$, almost surely. 

\end{theorem}

\begin{IEEEproof} Consider the Lyapunov-like function $V(\cdot)$
given by $V(x)\triangleq x^{\mathrm{T}}\hat{P}x$, $x\in\mathbb{R}^{n}$.
For the case $\tau_{i}=i$, $i\in\mathbb{N}_{0}$, by (\ref{eq:system}),
(\ref{eq:control-input}), we have 
\begin{align}
V(x(t+1)) & =x^{\mathrm{T}}(t)\left(A+\left(1-l(t)\right)BK\right)^{\mathrm{T}}\hat{P}\nonumber \\
 & \quad\,\cdot\left(A+\left(1-l(t)\right)BK\right)x(t),\,t\in\mathbb{N}_{0}.\label{eq:vattauplush-1}
\end{align}
 From (\ref{eq:betacond-1}), (\ref{eq:varphicond-1}), and (\ref{eq:vattauplush-1}),
this can be bounded by 
\begin{align}
V(x(t+1)) & \geq(1-l(t))\hat{\beta}V(x(t))+l(t)\hat{\varphi}V(x(t))\label{eq:vineq-1}
\end{align}
 for $t\in\mathbb{N}_{0}$. Now, let $\eta(k)\triangleq\prod_{t=0}^{k-1}\left((1-l(t))\hat{\beta}+l(t)\hat{\varphi}\right)$.
It follows from (\ref{eq:vineq-1}) that 
\begin{align}
V(x(k)) & \geq\eta(k)V(x(0))\label{eq:vetaineq-1}
\end{align}
for $k\in\mathbb{N}$. Furthermore, since $\ln\left[(1-q)\hat{\beta}+q\hat{\varphi}\right]=(1-q)\ln\hat{\beta}+q\ln\hat{\varphi}$
for $q\in\{0,1\}$, we have 
\begin{align*}
\ln\eta(k) & =\sum_{t=0}^{k-1}(1-l(t))\ln\hat{\beta}+\sum_{t=0}^{k-1}l(t)\ln\hat{\varphi}
\end{align*}
Now since $\hat{\beta}\in(0,1)$, we have $\ln\hat{\beta}<0$, and
hence by (\ref{eq:long-run-average-bound-1}), 
\begin{align*}
 & \liminf_{k\to\infty}\frac{1}{k}\sum_{t=0}^{k-1}(1-l(t))\ln\hat{\beta}=(\ln\hat{\beta})\limsup_{k\to\infty}\frac{1}{k}\sum_{t=0}^{k-1}(1-l(t))\\
 & \,\,=(\ln\hat{\beta})(1-\liminf_{k\to\infty}\frac{1}{k}\sum_{t=0}^{k-1}l(t))\geq(\ln\hat{\beta})(1-\sigma).
\end{align*}
Furthermore, since $\ln\hat{\varphi}\geq0$, it follows from (\ref{eq:long-run-average-bound-1})
that $\liminf_{k\to\infty}\frac{1}{k}\sum_{t=0}^{k-1}l(t)\ln\hat{\varphi}\geq\sigma\ln\hat{\varphi}$.
Consequently, by (\ref{eq:betaandvarphicond-1}), 
\begin{align*}
 & \liminf_{k\to\infty}\frac{\ln\eta(k)}{k}\geq\liminf_{k\to\infty}\frac{1}{k}\sum_{t=0}^{k-1}(1-l(t))\ln\hat{\beta}\\
 & \quad+\liminf_{k\to\infty}\frac{1}{k}\sum_{t=0}^{k-1}l(t)\ln\hat{\varphi}\geq(1-\sigma)\ln\hat{\beta}+\sigma\ln\hat{\varphi}>0,
\end{align*}
 almost surely. As a consequence, $\lim_{k\to\infty}\ln\eta(k)=\infty$,
and hence, $\lim_{k\to\infty}\eta(k)=\infty$, almost surely. Thus,
it follows from (\ref{eq:vetaineq-1}) that $\mathbb{P}[\lim_{t\to\infty}V(x(t))=\infty]=1$,
which implies that $\lim_{t\to\infty}\|x(t)\|=\infty$, almost surely.
\end{IEEEproof} \vskip 5pt

Theorem~\ref{TheoremMain-Attackers-Perspective} provides sufficient
conditions (\ref{eq:betacond-1})--(\ref{eq:betaandvarphicond-1})
to assess \emph{instability} of the closed-loop system (\ref{eq:system}),
(\ref{eq:control-input}). These conditions are complementary to the
\emph{stability} conditions (\ref{eq:betacond})--(\ref{eq:betaandvarphicond})
in Theorem~\ref{TheoremMain}. This point is further illustrated
by focusing on the scalar systems case. 

\begin{newexample}Consider the scalar system (\ref{eq:system}) with
$A,B\in\mathbb{R}$. Then, conditions (\ref{eq:betacond-1}), (\ref{eq:varphicond-1})
as well as (\ref{eq:betacond}), (\ref{eq:varphicond}) can be satisfied
by $\hat{P}=P=1$, $\hat{\beta}=\beta=(A+BK)^{2}$, and $\hat{\varphi}=\varphi=A^{2}$.
Now, if $\lim_{k\to\infty}\frac{L(k)}{k}$ exists and is a fixed constant,
we can set $\sigma=\rho=\lim_{k\to\infty}\frac{L(k)}{k}$ in (\ref{eq:betaandvarphicond})
and (\ref{eq:betaandvarphicond-1}) to obtain the stability condition
\begin{align}
(1-\lim_{k\to\infty}\frac{L(k)}{k})(A+BK)^{2}+\lim_{k\to\infty}\frac{L(k)}{k}A^{2} & <0,\label{eq:scalar-stability}
\end{align}
 and the instability condition 
\begin{align}
(1-\lim_{k\to\infty}\frac{L(k)}{k})(A+BK)^{2}+\lim_{k\to\infty}\frac{L(k)}{k}A^{2} & >0.\label{eq:scalar-instability}
\end{align}
 The limit $\lim_{k\to\infty}\frac{L(k)}{k}$ is a fixed constant
for example when the packet losses are Bernoulli-type or periodic.
In those cases, (\ref{eq:scalar-stability}) and (\ref{eq:scalar-instability})
indicate that Theorems~\ref{TheoremMain} and \ref{TheoremMain-Attackers-Perspective}
provide tight stability/instability conditions for scalar systems.
On the other hand, for multi-dimensional systems, scalars $\beta$
and $\hat{\beta}$ as well as $\varphi$ and $\hat{\varphi}$ may
not always be selected equal to obtain tight results. Furthermore,
under random packet losses and malicious attacks, $\lim_{k\to\infty}\frac{L(k)}{k}$
may not always exist and hence there may be a discrepancy between
$\rho$ and $\sigma$ in (\ref{eq:long-run-average-bound}) and (\ref{eq:long-run-average-bound-1}).
\end{newexample}

Proposition~\ref{PropositionIndependentAttackStrategy} below provides
a range of values for $\sigma$ that satisfy (\ref{eq:c1-1}) in the
case where the network faces random and malicious packet losses. 

\begin{proposition} \label{PropositionIndependentAttackStrategy}
Consider the packet exchange failure indicator process $\{l(t)\in\{0,1\}\}_{t\in\mathbb{N}_{0}}$
given by (\ref{eq:ldefinitionforcombinedcase}) where $\{l_{\mathrm{R}}(t)\in\{0,1\}\}_{t\in\mathbb{N}_{0}}$
and $\{l_{\mathrm{M}}(t)\in\{0,1\}\}_{t\in\mathbb{N}_{0}}$ are mutually
independent. Suppose there exists $\sigma_{\mathrm{M}}\in(0,1)$ such
that 
\begin{align}
\sum_{k=1}^{\infty}\mathbb{P}[\sum_{t=0}^{k-1}l_{\mathrm{M}}(t)<\sigma_{\mathrm{M}}k] & <\infty.\label{eq:reverse-jamming-condition}
\end{align}
Furthermore, suppose $\{l_{\mathrm{R}}(t)\in\{0,1\}\}_{t\in\mathbb{N}_{0}}$
satisfies (\ref{eq:p0condition-1}) with $p_{0}\in(0,1)$. Then (\ref{eq:c1-1})
holds for all $\sigma\in(0,1-p_{0}(1-\sigma_{\mathrm{M}}))$.

\end{proposition}\begin{IEEEproof} First, by using (\ref{eq:ldefinitionforcombinedcase}),
we obtain 
\begin{align}
 & \mathbb{P}[\sum_{t=0}^{k-1}l(t)<\sigma k]=\mathbb{P}[\sum_{t=0}^{k-1}(1-l(t))>(1-\sigma)k]\nonumber \\
 & \,\,=\mathbb{P}[\sum_{t=0}^{k-1}(1-l_{\mathrm{R}}(t))(1-l_{\mathrm{M}}(t))>(1-\sigma)k],\,\,k\in\mathbb{N}.\label{eq:sigmavarrhorelation}
\end{align}
Furthermore, it follows from (\ref{eq:reverse-jamming-condition})
that $\sum_{k=1}^{\infty}\mathbb{P}[\sum_{t=0}^{k-1}(1-l_{\mathrm{M}}(t))>(1-\sigma_{\mathrm{M}})k]=\sum_{k=1}^{\infty}\mathbb{P}[\sum_{t=0}^{k-1}l_{\mathrm{M}}(i)<\sigma_{\mathrm{M}}k]<\infty.$
Hence, $\{\chi(i)\triangleq\{0,1\}\}_{i\in\mathbb{N}_{0}}$ defined
by $\chi(i)=1-l_{\mathrm{\mathrm{M}}}(i),i\in\mathbb{N}_{0}$, satisfies
(\ref{eq:chicond}) with $\tilde{w}=1-\sigma_{\mathrm{M}}<1$. Furthermore,
$\{\xi(i)\triangleq\{0,1\}\}_{i\in\mathbb{N}_{0}}$ defined by $\xi(i)=1-l_{\mathrm{R}}(i),i\in\mathbb{N}_{0}$,
satisfies (\ref{eq:xicond}) with $\tilde{p}=p_{0}\in(0,1)$. We then
have from Lemma~\ref{KeyMarkovLemma} that 
\begin{align}
 & \sum_{k=1}^{\infty}\mathbb{P}[\sum_{t=0}^{k-1}(1-l_{\mathrm{R}}(t))(1-l_{\mathrm{M}}(t))>\varrho k]<\infty,\label{eq:oneminussigmainf}
\end{align}
 for all $\varrho\in(p_{0}(1-\sigma_{\mathrm{M}}),1-\sigma_{\mathrm{M}})$. 

In the rest of the proof, we will show that (\ref{eq:oneminussigmainf})
holds also for $\varrho\in[1-\sigma_{\mathrm{M}},1)$. To this end,
let $\varrho'\triangleq\frac{p_{0}(1-\sigma_{\mathrm{M}})+1-\sigma_{\mathrm{M}}}{2}$.
Since $\varrho'\in(p_{0}(1-\sigma_{\mathrm{M}}),1-\sigma_{\mathrm{M}})$,
by (\ref{eq:oneminussigmainf}), we get $\lambda\triangleq\sum_{k=1}^{\infty}\mathbb{P}[\sum_{t=0}^{k-1}(1-l_{\mathrm{R}}(t))(1-l_{\mathrm{M}}(t))>\varrho'k]<\infty$.
Furthermore, for all $\varrho\in[1-\sigma_{\mathrm{M}},1)$ we have
$\varrho\geq\varrho'$ and hence 
\begin{align*}
 & \mathbb{P}[\sum_{t=0}^{k-1}(1-l_{\mathrm{R}}(t))(1-l_{\mathrm{M}}(t))>\varrho k]\\
 & \quad\leq\mathbb{P}[\sum_{t=0}^{k-1}(1-l_{\mathrm{R}}(t))(1-l_{\mathrm{M}}(t))>\varrho'k],\quad k\in\mathbb{N}.
\end{align*}
Thus, for $\varrho\in[1-\sigma_{\mathrm{M}},1)$, $\sum_{k=1}^{\infty}\mathbb{P}[\sum_{t=0}^{k-1}(1-l_{\mathrm{R}}(t))(1-l_{\mathrm{M}}(t))>\varrho k]\leq\lambda<\infty$.
Therefore, (\ref{eq:oneminussigmainf}) holds for all $\varrho\in(p_{0}(1-\sigma_{\mathrm{M}}),1)=(p_{0}(1-\sigma_{\mathrm{M}}),1-\sigma_{\mathrm{M}})\cup[1-\sigma_{\mathrm{M}},1)$.
Now since $\sigma=1-\varrho$, it follows from (\ref{eq:sigmavarrhorelation})
that (\ref{eq:c1-1}) holds for all $\sigma\in(0,1-p_{0}(1-\sigma_{\mathrm{M}}))$.
\end{IEEEproof}

Proposition~\ref{PropositionIndependentAttackStrategy} shows that
when malicious attacks are independent of the random losses and they
satisfy (\ref{eq:reverse-jamming-condition}), the inequalities (\ref{eq:c1-1})
and (\ref{eq:long-run-average-bound-1}) (due to Lemma~ \ref{key-lemma1-1})
hold for a range of values of $\sigma$. This result indicates the
effects of independent random packet losses and malicious attacks
on the asymptotic ratio of packet exchange attempt failures over all
attempts. 

The next result is concerned with the scenarios where random packet
losses and malicious attacks need not be independent. 

\begin{proposition} \label{PropositionDependentAttackStrategy-1}
Consider the packet exchange failure indicator process $\{l(t)\in\{0,1\}\}_{t\in\mathbb{N}_{0}}$
given by (\ref{eq:ldefinitionforcombinedcase}). Suppose there exists
$\sigma_{\mathrm{M}}\in(0,1)$ such that (\ref{eq:reverse-jamming-condition})
holds. Furthermore, suppose $\{l_{\mathrm{R}}(t)\in\{0,1\}\}_{t\in\mathbb{N}_{0}}$
satisfies (\ref{eq:p0condition-1}) with $p_{0}\in(0,1)$. Then (\ref{eq:c1-1})
holds for all $\sigma\in(0,\max\{1-p_{0},\sigma_{\mathrm{M}}\})$.

\end{proposition}\begin{IEEEproof} We will show that (\ref{eq:c1-1})
holds for the cases: 1) $\max\{1-p_{0},\sigma_{\mathrm{M}}\}=1-p_{0}$
and 2) $\max\{1-p_{0},\sigma_{\mathrm{M}}\}=\sigma_{\mathrm{M}}$.
First, if $\max\{1-p_{0},\sigma_{\mathrm{M}}\}=1-p_{0}$, then noting
that $\sum_{t=0}^{k-1}l_{\mathrm{R}}(t)\leq\sum_{t=0}^{k-1}l(t)$,
we obtain 
\begin{align}
\mathbb{P}[\sum_{t=0}^{k-1}l(t)<\sigma k] & =\mathbb{P}[\sum_{t=0}^{k-1}(1-l(t))>(1-\sigma)k]\nonumber \\
 & \leq\mathbb{P}[\sum_{t=0}^{k-1}(1-l_{\mathrm{R}}(t))>(1-\sigma)k],\label{eq:lrineqp0}
\end{align}
for $k\in\mathbb{N}$. Now, $\{\chi(i)\triangleq\{0,1\}\}_{i\in\mathbb{N}_{0}}$
with $\chi(i)=1,i\in\mathbb{N}_{0}$, satisfies (\ref{eq:chicond})
with $\tilde{w}=1$. Furthermore, $\{\xi(i)\triangleq\{0,1\}\}_{i\in\mathbb{N}_{0}}$
with $\xi(i)=1-l_{\mathrm{R}}(i),i\in\mathbb{N}_{0}$, satisfies (\ref{eq:xicond})
with $\tilde{p}=p_{0}\in(0,1)$. Since $1-\sigma>p_{0}$, we have
from Lemma~\ref{KeyMarkovLemma} that $\sum_{k=1}^{\infty}\mathbb{P}[\sum_{t=0}^{k-1}(1-l_{\mathrm{R}}(t))>(1-\sigma)k]<\infty$.
Hence, by (\ref{eq:lrineqp0}), we have (\ref{eq:c1-1}). 

Next, if $\max\{1-p_{0},\sigma_{\mathrm{M}}\}=\sigma_{\mathrm{M}}$,
then since $\sum_{t=0}^{k-1}l_{\mathrm{M}}(t)\leq\sum_{t=0}^{k-1}l(t)$
and $\sigma<\sigma_{\mathrm{M}}$, we get 
\begin{align}
\mathbb{P}[\sum_{t=0}^{k-1}l(t)<\sigma k] & \leq\mathbb{P}[\sum_{t=0}^{k-1}l_{\mathrm{M}}(t)<\sigma_{\mathrm{M}}k],\quad k\in\mathbb{N}.\label{eq:lmineqsigmam}
\end{align}
 Consequently, (\ref{eq:reverse-jamming-condition}) and (\ref{eq:lmineqsigmam})
imply (\ref{eq:c1-1}).\end{IEEEproof}

Proposition~\ref{PropositionDependentAttackStrategy-1} provides
a range for $\sigma$ in (\ref{eq:c1-1}) when we consider the case
where random packet losses and malicious attacks may be dependent.
This range is smaller in comparison to the one provided in Proposition~\ref{PropositionIndependentAttackStrategy}
for the independent case. This is because Proposition~\ref{PropositionDependentAttackStrategy-1}
deals with scenarios including \emph{the worst case from the perspective
of the attacker}. In that scenario, the malicious attacks and random
packet losses happen at the same time instants, and hence, the statistical
frequency of the overall packet exchange failures cannot exceed the
maximum of the frequencies of malicious attacks and random packet
losses. We remark that there are other scenarios where the attacks
depend on the random packet losses. For instance, the attacker may
intentionally avoid attacking when there is already a random packet
loss. This scenario is characterized in the mathematical setting by
$l_{\mathrm{R}}(t)l_{\mathrm{M}}(t)=0$, $t\in\mathbb{N}_{0}$. For
this scenario, the following proposition provides a range of $\sigma$
that satisfy (\ref{eq:c1-1}). 

\begin{proposition} \label{PropositionDependentAttackStrategy-2}
Consider the packet exchange failure indicator process $\{l(t)\in\{0,1\}\}_{t\in\mathbb{N}_{0}}$
given by (\ref{eq:ldefinitionforcombinedcase}). Suppose $\{l_{\mathrm{R}}(t)\in\{0,1\}\}_{t\in\mathbb{N}_{0}}$
satisfies (\ref{eq:p0condition-1}) with $p_{0}\in(0,1)$. Furthermore,
suppose $l_{\mathrm{R}}(t)l_{\mathrm{M}}(t)=0$, $t\in\mathbb{N}_{0}$,
and there exists $\sigma_{\mathrm{M}}\in(0,1)$ such that (\ref{eq:reverse-jamming-condition})
holds. If $1-p_{0}+\sigma_{\mathrm{M}}\leq1$, then (\ref{eq:c1-1})
holds for all $\sigma\in(0,1-p_{0}+\sigma_{\mathrm{M}})$.

\end{proposition}\begin{IEEEproof} First let $\epsilon\triangleq1-p_{0}+\sigma_{\mathrm{M}}-\sigma$,
and define $\sigma_{1}\triangleq\max\{0,1-p_{0}-\frac{\epsilon}{2}\}$,
$\sigma_{2}\triangleq\max\{0,\sigma_{\mathrm{M}}-\frac{\epsilon}{2}\}$.
Note that $\sigma_{1}+\sigma_{2}\geq\sigma$. Now since $l_{\mathrm{R}}(t)l_{\mathrm{M}}(t)=0$,
$t\in\mathbb{N}_{0}$, we have from (\ref{eq:ldefinitionforcombinedcase})
that $\sum_{t=0}^{k-1}l(t)=\sum_{t=0}^{k-1}l_{\mathrm{R}}(t)+\sum_{t=0}^{k-1}l_{\mathrm{M}}(t)$.
As a result 
\begin{align}
 & \mathbb{P}[\sum_{t=0}^{k-1}l(t)<\sigma k]=\mathbb{P}[\sum_{t=0}^{k-1}l_{\mathrm{R}}(t)+\sum_{t=0}^{k-1}l_{\mathrm{M}}(t)<\sigma k]\nonumber \\
 & \,\,\leq\mathbb{P}[\sum_{t=0}^{k-1}l_{\mathrm{R}}(t)+\sum_{t=0}^{k-1}l_{\mathrm{M}}(t)<\sigma_{1}k+\sigma_{2}k]\nonumber \\
 & \,\,\leq\mathbb{P}[\sum_{t=0}^{k-1}l_{\mathrm{R}}(t)<\sigma_{1}k]+\mathbb{P}[\sum_{t=0}^{k-1}l_{\mathrm{M}}(t)<\sigma_{2}k],\,\,k\in\mathbb{N}.\label{eq:dep2sum}
\end{align}
If $\sigma_{1}=0$, then $\mathbb{P}[\sum_{t=0}^{k-1}l_{\mathrm{R}}(t)<\sigma_{1}k]=0$,
and hence $\sum_{k=1}^{\infty}\mathbb{P}[\sum_{t=0}^{k-1}l_{\mathrm{R}}(t)<\sigma_{1}k]=0<\infty$.
If, on the other hand, $\sigma_{1}>0$, then we can utilize Lemma~\ref{KeyMarkovLemma}.
Specifically, $\{\chi(i)\triangleq\{0,1\}\}_{i\in\mathbb{N}_{0}}$
with $\chi(i)=1,i\in\mathbb{N}_{0}$, satisfies (\ref{eq:chicond})
with $\tilde{w}=1$. Furthermore, $\{\xi(i)\triangleq\{0,1\}\}_{i\in\mathbb{N}_{0}}$
with $\xi(i)=1-l_{\mathrm{R}}(i),i\in\mathbb{N}_{0}$, satisfies (\ref{eq:xicond})
with $\tilde{p}=p_{0}\in(0,1)$. Since $\sigma_{1}>0$, it means that
$\sigma_{1}=1-p_{0}-\frac{\epsilon}{2}$. Now, since $\epsilon>0$,
we have $(1-\sigma_{1})\in(p_{0},1)$. Consequently, we obtain from
Lemma~\ref{KeyMarkovLemma} that $\sum_{k=1}^{\infty}\mathbb{P}[\sum_{t=0}^{k-1}(1-l_{\mathrm{R}}(t))>(1-\sigma_{1})k]<\infty$,
and hence, 
\begin{align}
 & \sum_{k=1}^{\infty}\mathbb{P}[\sum_{t=0}^{k-1}l_{\mathrm{R}}(t)<\sigma_{1}k]\nonumber \\
 & \quad=\sum_{k=1}^{\infty}\mathbb{P}[\sum_{t=0}^{k-1}(1-l_{\mathrm{R}}(t))>(1-\sigma_{1})k]<\infty.\label{eq:firstsumdep2}
\end{align}
 Similarly, if $\sigma_{2}=0$, then $\mathbb{P}[\sum_{t=0}^{k-1}l_{\mathrm{M}}(t)<\sigma_{2}k]=0$,
and hence $\sum_{k=1}^{\infty}\mathbb{P}[\sum_{t=0}^{k-1}l_{\mathrm{M}}(t)<\sigma_{2}k]=0<\infty$.
On the other hand, if $\sigma_{2}=\sigma_{\mathrm{M}}-\frac{\epsilon}{2}>0$,
since $\epsilon>0$, we have $\sigma_{2}<\sigma_{\mathrm{M}}$. Thus,
$\mathbb{P}[\sum_{t=0}^{k-1}l_{\mathrm{M}}(t)<\sigma_{2}k]\leq\mathbb{P}[\sum_{t=0}^{k-1}l_{\mathrm{M}}(t)<\sigma_{\mathrm{M}}k]$.
It then follows from (\ref{eq:reverse-jamming-condition}) that 
\begin{align}
 & \sum_{k=1}^{\infty}\mathbb{P}[\sum_{t=0}^{k-1}l_{\mathrm{M}}(t)<\sigma_{2}k]\leq\sum_{k=1}^{\infty}\mathbb{P}[\sum_{t=0}^{k-1}l_{\mathrm{M}}(t)<\sigma_{\mathrm{M}}k]<\infty.\label{eq:secondsumdep2}
\end{align}
 Finally, (\ref{eq:c1-1}) follows from (\ref{eq:dep2sum})--(\ref{eq:secondsumdep2}).\end{IEEEproof}

An attacker that is knowledgeable about the random packet losses in
the network may avoid placing malicious attacks when random packet
losses occur. Proposition~\ref{PropositionDependentAttackStrategy-2}
provides a range of values of $\sigma$ such that the inequality (\ref{eq:c1-1})
holds when the attacker follows this strategy. Compared to the case
where attacks and random packet losses are independent, this strategy
would increase the overall number of packet exchange failures, even
though the number of attacks may be the same. The reason is that in
the independent case, the attacks and random packet losses may occasionally
happen at the same time, reducing the total packet failure count.
Noe that the range of $\sigma$ in Proposition~\ref{PropositionDependentAttackStrategy-2}
is larger than that in Proposition~\ref{PropositionIndependentAttackStrategy},
where the attacks and random packet losses are independent, even though
in both results the malicious attacks satisfy (\ref{eq:reverse-jamming-condition})
with the same $\sigma_{\mathrm{M}}\in(0,1)$. In Section~\ref{sec:Numerical-Example}-B,
we discuss and compare two attack strategies independent/dependent
on random packet losses. Both strategies cause instability for certain
feedback gain and event-triggering mechanism parameters. 

It is important to note that particular choices of the controller
parameters may result in instability when $\sigma\in[0,1]$ is large.
If the packet exchange failures are known to happen statistically
frequently, that is, if $\sigma$ is large, then the feedback gain
$K$ and the event-triggering mechanism parameters $\beta$ and $P$
should be redesigned to ensure stability. In such cases, Theorem~\ref{TheoremMain}
and Corollary~\ref{Corollary} can be employed with $\rho\geq\sigma$
that satisfies (\ref{eq:lcond1}) or (\ref{eq:long-run-average-bound}).

\section{Numerical Examples \label{sec:Numerical-Example}}

In this section we present numerical examples to illustrate our results
provided in Sections~\ref{sec:Network-Control-Problem}--\ref{sec:Attacker's-Perspective}.

\subsubsection*{A) Example 1}

We consider the system (\ref{eq:system}) with 
\begin{align*}
A\triangleq\left[\begin{array}{cc}
1 & 0.1\\
-0.5 & 1.1
\end{array}\right], & \quad B\triangleq\left[\begin{array}{c}
0.1\\
1.2
\end{array}\right].
\end{align*}
 We use the event-triggering control law (\ref{eq:control-input}),
(\ref{eq:attemptedpacketexchangetimes}) for stabilization of (\ref{eq:system})
over a network that faces independent random packet losses and malicious
attacks. Specifically, random packet losses are assumed to be characterized
by the Markov chain $\{l_{\mathrm{R}}(i)\in\{0,1\}\}_{i\in\mathbb{N}_{0}}$
with initial distribution $\vartheta_{0}=0$, $\vartheta_{1}=1$,
and transition probabilities $p_{0,1}(i)\triangleq0.2+0.03\sin^{2}(0.1i),$
$p_{1,1}(i)\triangleq0.2+0.03\cos^{2}(0.1i)$, and $p_{q,0}(i)=1-p_{q,1}(i)$,
$q\in\{0,1\}$, $i\in\mathbb{N}_{0}$. Note that $\{l_{\mathrm{R}}(i)\in\{0,1\}\}_{i\in\mathbb{N}_{0}}$
satisfies (\ref{eq:p1condition-1}) and (\ref{eq:p0condition-1})
with $p_{1}=0.23$ and $p_{0}=0.8$. Furthermore, the network is subject
to jamming attacks that is independent of $\{l_{\mathrm{R}}(i)\in\{0,1\}\}_{i\in\mathbb{N}_{0}}$
and satisfies (\ref{eq:jammingcondition}) with $\kappa=2$ and $\tau=5$.
By Lemma~\ref{JammingLemma}, (\ref{eq:general-attack-condition})
holds with $\rho_{\mathrm{M}}=0.21$ since $\rho_{\mathrm{M}}=0.21>\frac{1}{\tau}=0.2$.
Furthermore, note that $p_{1}+p_{0}\rho_{\mathrm{M}}<0.4$. Hence,
it follows from Proposition~\ref{PropositionCombinedCase} that for
$\rho=0.4$, (\ref{eq:lcond1}) of Assumption~\ref{MainAssumption}
holds, which implies (\ref{eq:long-run-average-bound}) through Lemma~\ref{key-lemma1}. 

We designed the controller based on the procedure in Section~\ref{sub:Feedback-Gain-Design}
and obtained the matrices 
\begin{align*}
Q & =\left[\begin{array}{cc}
0.618 & -2.119\\
-2.119 & 28.214
\end{array}\right],\,\,M=\left[\begin{array}{cc}
0.202 & -20.405\end{array}\right],
\end{align*}
 and scalars $\beta=0.55$, $\varphi=2.4516$ satisfy (\ref{eq:corolcond1}),
(\ref{eq:corolcond2}), and (\ref{eq:betaandvarphicond}) with $\rho=0.4$.
Hence, it follows from Corollary~\ref{Corollary} that the event-triggered
control law (\ref{eq:control-input}), (\ref{eq:attemptedpacketexchangetimes})
with $P=Q^{-1}$ and $K=MQ^{-1}$ guarantees almost sure asymptotic
stabilization.

\begin{figure}[t]
\begin{center}\includegraphics[width=0.87\columnwidth]{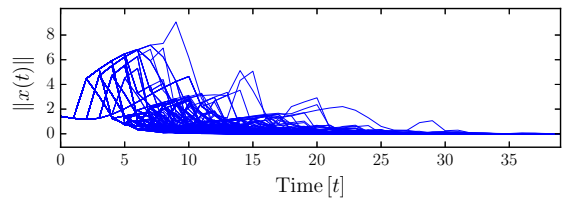}\end{center}\vskip -20pt\caption{Sample paths of the state norm}
 \vskip -15pt \label{Flo:statenorm}
\end{figure}

\begin{figure}[t]
\begin{center}\includegraphics[width=0.87\columnwidth]{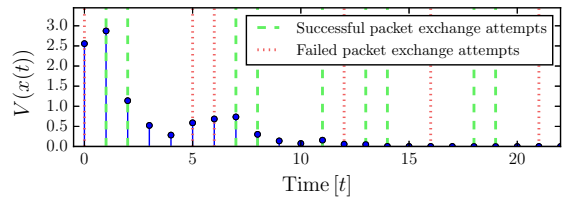}\end{center}\vskip -20pt\caption{A sample path of Lyapunov-like function $V(\cdot)$}
 \vskip -10pt \label{Flo:v1}
\end{figure}

We generated $250$ sample state trajectories using the same initial
condition $x_{0}=\left[1,\,1\right]^{\mathrm{T}}$ and the event-triggering
mechanism parameter $\theta=1000$, but with different sample paths
for $l_{\mathrm{R}}(\cdot)$ and $l_{\mathrm{M}}(\cdot)$. We can
check in Fig.~\ref{Flo:statenorm} that all state trajectories go
to the origin. The same is true for the Lyapunov-like function $V(x(t))$.
We show a single sample trajectory of $V(\cdot)$ in Fig.~\ref{Flo:v1}.
The Lyapunov-like function $V(\cdot)$ converges to zero, but notice
that it is not monotonically decreasing. The Lyapunov-like function
$V(\cdot)$ increases in two situations. First, when packet exchange
attempts fail, $V(\cdot)$ may grow and take a larger value at the
next packet exchange attempt instant due to unstable dynamics of the
uncontrolled system. Second, $V(\cdot)$ may also increase some time
after a successful packet exchange between the plant and the controller.
This is because the constant control input updated with the packet
exchange becomes ineffective after some time. Note that eventually
a new packet exchange attempt is triggered before $V(\cdot)$ leaves
the bound identified in the event-triggering condition (\ref{eq:attemptedpacketexchangetimes}).

\subsubsection*{B) Example 2}

\label{sub:AttackExample}

Our goal in this example is to illustrate effects of different attack
strategies discussed in Section~\ref{sec:Attacker's-Perspective}.
Here, we consider a scalar linear system (\ref{eq:system}) with $A=2$
and $B=1$. Its initial state is set to $x_{0}=1$. Furthermore, the
feedback gain and the event-triggering mechanism parameters in (\ref{eq:control-input})
and (\ref{eq:attemptedpacketexchangetimes}) are given by $K=-1.75$,
$\beta=0.0625$, $P=1$. We set the packet exchange events to be triggered
at all time instants. This is done with $\theta=1$ in (\ref{eq:attemptedpacketexchangetimes}). 

The random packet losses in the network are characterized by the Markov
chain $\{l_{\mathrm{R}}(t)\in\{0,1\}\}_{t\in\mathbb{N}_{0}}$ with
initial distribution $\vartheta_{0}=0$, $\vartheta_{1}=1$, and transition
probabilities $p_{0,1}(t)\triangleq0.4+0.01\cos(0.1t),$ $p_{1,1}(t)\triangleq0.4+0.01\sin(0.1t)$,
and $p_{q,0}(t)=1-p_{q,1}(t)$, $q\in\{0,1\}$, $t\in\mathbb{N}_{0}$.
Note that $l_{\mathrm{R}}(\cdot)$ satisfies (\ref{eq:p1condition-1})
and (\ref{eq:p0condition-1}) with $p_{1}=0.41$ and $p_{0}=0.61$. 

We consider two attack strategies described by (\ref{eq:jammingcondition})
and discuss stability properties of the closed-loop system. 

\begin{figure}[t]
\begin{center}\includegraphics[width=0.9\columnwidth]{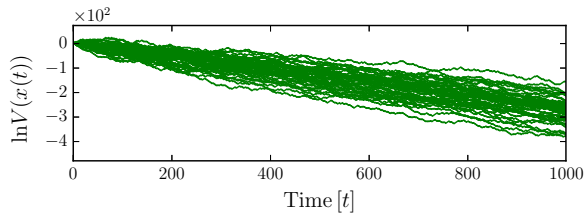}\end{center}\vskip -20pt\caption{Sample paths of $\ln V(x(\cdot))$ under malicious attack (\ref{eq:strategy-one})
with $\tau=3$}
 \vskip -10pt \label{Flo:lnvcase1tau3}
\end{figure}

\begin{figure}[t]
\begin{center}\includegraphics[width=0.9\columnwidth]{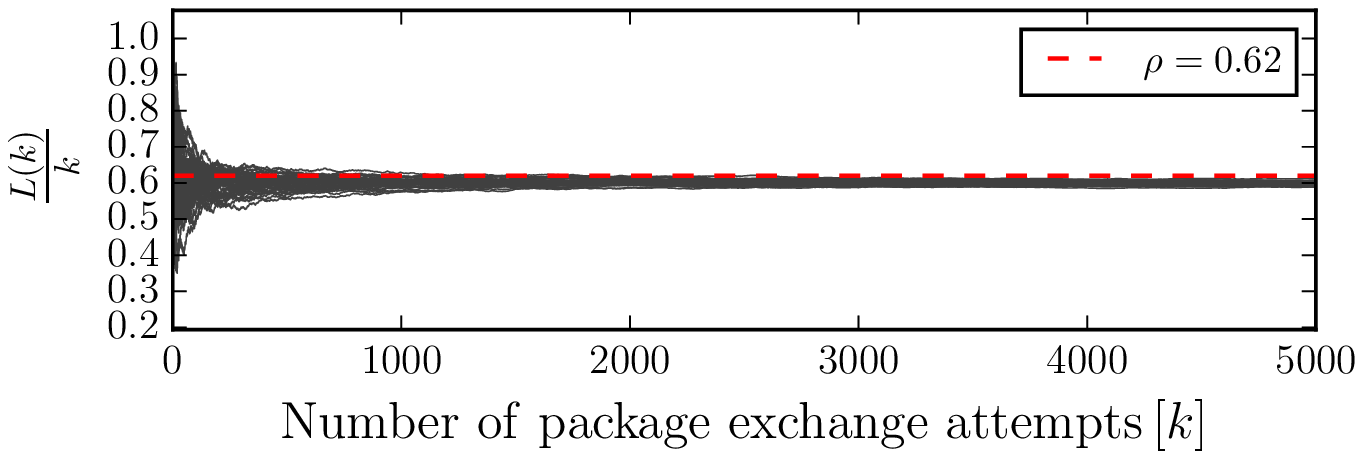}\end{center}\vskip -20pt\caption{Sample paths of the average number of packet exchange attempt failures
($L(k)/k$) under malicious attack (\ref{eq:strategy-one}) with $\tau=3$}
\vskip -10pt \label{Flo:lcase1tau3}
\end{figure}

\emph{(i) Random-Loss-Independent Attack Strategy:} We consider the
strategy given by $l_{\mathrm{M}}(0)\triangleq0$, and 
\begin{align}
l_{\mathrm{M}}(t)\triangleq\begin{cases}
1, & \mathrm{if}\,\,\sum_{i=0}^{t-1}l_{\mathrm{M}}(i)\leq\frac{t+1}{\tau}-1,\\
0, & \mathrm{otherwise},
\end{cases}\quad & t\in\mathbb{N}.\label{eq:strategy-one}
\end{align}
 Note that (\ref{eq:strategy-one}) satisfies (\ref{eq:jammingcondition})
with $\kappa=0$. In this strategy, the attacker uses the total count
of all attacks prior to time $t$ to check whether placing an attack
at time $t$ would meet the requirement in (\ref{eq:jammingcondition})
or not. The attacker causes a packet exchange failure at time $t$
if (\ref{eq:jammingcondition}) still holds at time $t$ (i.e., $\sum_{i=0}^{t}l_{\mathrm{M}}(i)\leq\frac{t+1}{\tau}$).
Under this strategy, attacks are independent of random packet losses
and the attack times become periodic with period $\tau$ when $\tau$
is an integer. We will assess stability/instability of the closed-loop
system with two different values of $\tau$. 

\begin{figure}[t]
\begin{center}\includegraphics[width=0.9\columnwidth]{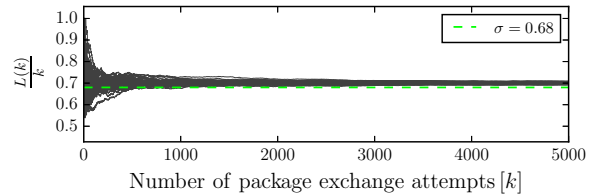}\end{center}\vskip -20pt\caption{Sample paths of the average number of packet exchange attempt failures
($L(k)/k$) under malicious attack (\ref{eq:strategy-one}) with $\tau=2$}
\vskip -10pt \label{Flo:lcase1tau2}
\end{figure}

First, we consider $\tau=3$, that is, the attacker prevents packet
exchanges once in every $3$ steps. In this case the closed-loop system
is stable despite the attack. We use Theorem~\ref{TheoremMain} to
show stability as follows. By Lemma~\ref{JammingLemma}, (\ref{eq:general-attack-condition})
holds with $\rho_{\mathrm{M}}=0.3334>\frac{1}{\tau}=\frac{1}{3}$.
Now, note that $p_{1}+p_{0}\rho_{\mathrm{M}}<0.62$. Since $l_{\mathrm{R}}(\cdot)$
and $l_{\mathrm{M}}(\cdot)$ are independent, it follows from Proposition~\ref{PropositionCombinedCase}
that (\ref{eq:lcond1}) in Assumption~\ref{MainAssumption} holds
for $\rho=0.62$, which implies (\ref{eq:long-run-average-bound})
through Lemma~\ref{key-lemma1}. Further, (\ref{eq:betacond})--(\ref{eq:betaandvarphicond})
hold with $P=1$, $\beta=0.0625$, and $\varphi=4$. By Theorem~\ref{TheoremMain},
the event-triggered control law (\ref{eq:control-input}), (\ref{eq:attemptedpacketexchangetimes})
with $P=1$, $\beta=0.0625$, and $K=-1.75$ guarantees \emph{almost
sure asymptotic stabilization}. 

Fig.~\ref{Flo:lnvcase1tau3} shows $50$ sample trajectories of $\ln V(x(\cdot))$
where $V(x(t))\triangleq x^{2}(t)$. These trajectories are obtained
under malicious attack (\ref{eq:strategy-one}) but with different
sample paths for $\{l_{\mathrm{R}}(t)\in\{0,1\}\}_{t\in\mathbb{N}_{0}}$.
Note that all trajectories of $\ln V(x(\cdot))$ approach $-\infty$,
indicating convergence of the state to $0$. Moreover, in Fig.~\ref{Flo:lcase1tau3},
we show sample trajectories of the average number of packet exchange
attempt failures. Observe that the long run average number of packet
failures is small enough to guarantee stability ($\limsup_{k\to\infty}\frac{L(k)}{k}\leq\rho=0.62$). 

Next, we consider (\ref{eq:strategy-one}) with $\tau=2$, i.e., the
malicious attacker prevents every other packet exchange attempt. With
$\tau=2$, the closed-loop system becomes \emph{unstable}. We can
show this through Theorem~\ref{TheoremMain-Attackers-Perspective}
as follows. First, note that in this case, (\ref{eq:strategy-one})
implies (\ref{eq:reverse-jamming-condition}) with $\sigma_{\mathrm{M}}\triangleq0.49$.
To see this, we observe that $\sum_{t=0}^{k-1}l_{\mathrm{M}}(t)\geq\frac{k}{\tau}-1=\frac{k}{2}-1$
for all $k\in\mathbb{N}$. Next, using Markov's inequality we obtain
\begin{align*}
 & \mathbb{P}[\sum_{t=0}^{k-1}l_{\mathrm{M}}(t)<\sigma_{\mathrm{M}}k]=\mathbb{P}[\sum_{t=0}^{k-1}(1-l_{\mathrm{M}}(t))>(1-\sigma_{\mathrm{M}})k]\\
 & \quad\leq\mathbb{P}[e^{\sum_{t=0}^{k-1}(1-l_{\mathrm{M}}(t))}\geq e^{(1-\sigma_{\mathrm{M}})k}]\\
 & \quad\leq e^{-(1-\sigma_{\mathrm{M}})k}\mathbb{E}[e^{\sum_{t=0}^{k-1}(1-l_{\mathrm{M}}(t))}]\leq e^{-(1-\sigma_{\mathrm{M}})k}e^{1+\frac{k}{2}},
\end{align*}
 for $k\in\mathbb{N}$. Consequently, since $\sigma_{\mathrm{M}}<\frac{1}{\tau}=\frac{1}{2}$,
we have $\sum_{k=1}^{\infty}\mathbb{P}[\sum_{t=0}^{k-1}l_{\mathrm{M}}(t)<\sigma_{\mathrm{M}}k]\leq\sum_{k=1}^{\infty}e^{-(1-\sigma_{\mathrm{M}})k}e^{1+\frac{k}{2}}=e^{\frac{1}{2}+\sigma_{\mathrm{M}}}(1-e^{\sigma_{\mathrm{M}}-\frac{1}{2}})^{-1}<\infty,$
which implies (\ref{eq:reverse-jamming-condition}). Now, note that
$1-p_{0}(1-\sigma_{\mathrm{M}})>0.68$. Hence, by Proposition~\ref{PropositionIndependentAttackStrategy},
we have (\ref{eq:c1-1}) for $\sigma=0.68$. Consequently, by Lemma~\ref{key-lemma1-1},
(\ref{eq:long-run-average-bound-1}) holds for $\sigma=0.68$. Furthermore,
inequalities (\ref{eq:betacond-1})--(\ref{eq:betaandvarphicond-1})
hold with $\hat{P}=1$, $\hat{\beta}=0.0625$, and $\hat{\varphi}=4$.
It follows from Theorem~\ref{TheoremMain-Attackers-Perspective}
that the closed-loop system with $K=-1.75$ is \emph{unstable}. 

This example shows that an attacker can destabilize the system by
reducing $\tau$ from $3$ to $2$ and hence causing higher number
of packet losses on average. As illustrated in Fig.~\ref{Flo:lcase1tau2},
when $\tau=2$, in the long run, the average number of packet failures
becomes larger ($\liminf_{k\to\infty}\frac{L(k)}{k}\geq\sigma=0.68$)
compared to the case with $\tau=3$ in Fig.~\ref{Flo:lcase1tau3}. 

\emph{(ii) Selective Attack Strategy: }Next, we consider the case
where the attacker is knowledgeable about the random packet losses
in the network. To describe this strategy we let 
\begin{eqnarray}
\,l_{\mathrm{M}}(t)\triangleq\begin{cases}
1, & \mathrm{if}\,\,l_{\mathrm{R}}(t)=0\,\,\mathrm{and}\,\,\sum_{i=0}^{t-1}l_{\mathrm{M}}(i)\leq\frac{t+1}{\tau}-1,\\
0, & \mathrm{otherwise},
\end{cases}\label{eq:strategy-two}
\end{eqnarray}
 for $t\in\mathbb{N}$ and $l_{\mathrm{M}}(0)=0$. This strategy is
similar to the one given by (\ref{eq:strategy-one}) in that it satisfies
(\ref{eq:jammingcondition}) with $i=t$ and $\kappa=0$. However,
an attacker following (\ref{eq:strategy-two}) utilizes random packet
loss information at time $t$, by not placing an attack when $l_{\mathrm{R}}(t)=1$
(indicating packet failures due to random errors). 

\begin{figure}[t]
\begin{center}\includegraphics[width=0.95\columnwidth]{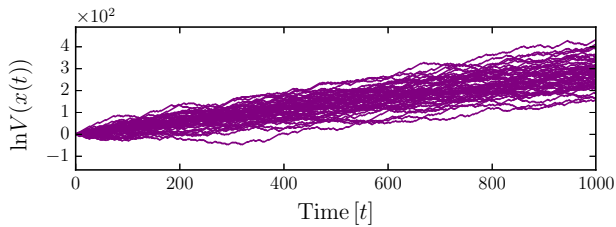}\end{center}\vskip -20pt\caption{Sample paths of $\ln V(x(\cdot))$ under malicious attack (\ref{eq:strategy-two})
with $\tau=3$}
\vskip -12pt \label{Flo:lnvcase2tauk175}
\end{figure}

\begin{figure}[t]
\begin{center}\includegraphics[width=0.95\columnwidth]{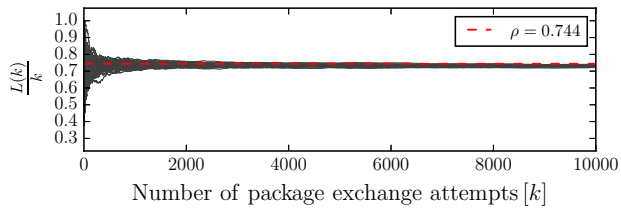}\end{center}\vskip -20pt\caption{Sample paths of the average number of packet exchange attempt failures
($L(k)/k$) under malicious attack (\ref{eq:strategy-two}) with $\tau=3$}
\vskip -12pt \label{Flo:lcase2}
\end{figure}

To compare, we set $\tau=3$, under which the first strategy (\ref{eq:strategy-one})
cannot destabilize the system. From the simulations, we notice that
with $\tau=3$, the selective attack strategy causes the system state
to diverge (see Fig.~\ref{Flo:lnvcase2tauk175} for $50$ sample
paths of $\ln V(x(t))$ with $V(x(t))\triangleq x^{2}(t)$). In fact,
we observe from Fig.~\ref{Flo:lcase2} that (\ref{eq:long-run-average-bound-1})
holds with $\sigma=0.7$. This $\sigma$ satisfies condition (\ref{eq:betaandvarphicond-1})
of Theorem~\ref{TheoremMain-Attackers-Perspective}, indicating instability. 

\begin{figure}[t]
\begin{center}\includegraphics[width=0.95\columnwidth]{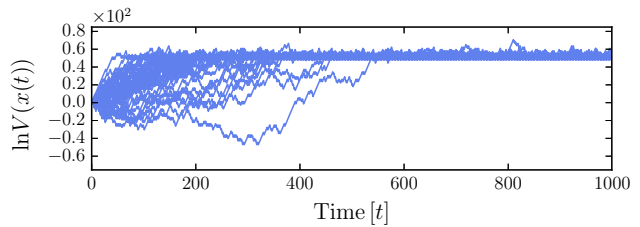}\end{center}\vskip -20pt\caption{Sample paths of $\ln V(x(\cdot))$ under state-dependent attacks }
\vskip -16pt \label{Flo:lnvcase2tauk175-statedependent}
\end{figure}

Next, we consider an extension to the attack model in (\ref{eq:strategy-two}).
In this model, the attacker places attacks whenever $l_{\mathrm{R}}(t)=0$,
$\sum_{i=0}^{t-1}l_{\mathrm{M}}(i)\leq\frac{t+1}{\tau}-1$, and $\ln V(x(t))\leq\zeta$.
From the simulations with $\zeta=50$ and $\tau=3$, we see that the
state does not diverge, but the attacker is able to keep it around
the level identified with $\ln V(x(t))=\zeta$ (see Fig. \ref{Flo:lnvcase2tauk175-statedependent}).
We also observe that in the long run, the average number of packet
failures approaches $\frac{2}{3}$. We remark that $\frac{2}{3}$
is a critical value for this example in the sense that $\rho<\frac{2}{3}$
in (\ref{eq:betaandvarphicond}) implies convergence of the state,
and $\sigma>\frac{2}{3}$ in (\ref{eq:betaandvarphicond-1}) implies
divergence. 

Finally, we show that by redesigning the feedback gain $K$, we can
reensure closed-loop system stability. To this end we set $K=-1.9$.
It follows from Theorem~\ref{TheoremMain} that the event-triggered
control law (\ref{eq:control-input}), (\ref{eq:attemptedpacketexchangetimes})
with $P=1$, $\beta=0.01$, guarantees \emph{almost sure asymptotic
stability} of the closed-loop system. This can be checked as follows.
By Lemma~\ref{JammingLemma}, the attack strategy (\ref{eq:strategy-two})
and its extension above satisfy (\ref{eq:general-attack-condition})
with $\rho_{\mathrm{M}}=0.3334$. Now, note that $p_{1}+\rho_{\mathrm{M}}<0.744$.
Since $l_{\mathrm{R}}(\cdot)$ and $l_{\mathrm{M}}(\cdot)$ are \emph{not
}independent, it follows from Proposition~\ref{PropositionDependentCombinedCase}
that (\ref{eq:lcond1}) of Assumption~\ref{MainAssumption} holds
for $\rho=0.744$, which implies (\ref{eq:long-run-average-bound})
through Lemma~\ref{key-lemma1}. Notice that Proposition~\ref{PropositionDependentCombinedCase}
provides a tight bound ($\rho=0.744$) for the average number of packet
exchange attempt failures for the attack strategy (\ref{eq:strategy-two})
(Fig.~\ref{Flo:lcase2}). Moreover, (\ref{eq:betacond})--(\ref{eq:betaandvarphicond})
hold with $P=1$, $\beta=0.01$, and $\varphi=4$.

\section{Conclusion }

In this paper, we explored control of linear dynamical systems over
networks that face random packet losses and malicious attacks. We
proposed a probabilistic characterization of the evolution of the
total number of packet exchange failures. Based on this characterization,
we obtained sufficient conditions for almost sure asymptotic stabilization
and presented a method for finding a stabilizing feedback gain and
parameters for our proposed event-triggered control framework. Furthermore,
to investigate potential cyber risks in networked control operations,
we studied the problem from the perspective of an attacker. We obtained
conditions under which combined effects of random and malicious packet
losses can destabilize the closed-loop system. 

The framework developed in this paper has been utilized to investigate
the \emph{output feedback control} problem in \cite{cetinkaya2015output}.
The probabilistic characterization developed in this paper is utilized
there for modeling random and malicious packet losses in transmission
of the output information from the plant sensors to the estimator
in the controller side. 

A direction for future research is to explore the networked control
problem when wireless communication is used. There, several communication
nodes and routers can be involved, and some of them may be compromised
by adversaries. Our proposed network model can be incorporated to
describe random failures and malicious attacks observed in such problems
\cite{ahmetnecsys2016}. Furthermore, there are other important issues
discussed in the networked control literature such as system and measurement
noise \cite{hespanha2007}, transmission delays \cite{schenato2007,donkers2011},
and the modeling of the communication protocol \cite{schenato2007,smarra2015}.
Investigation of these issues within our framework remains as a future
work. 

\bibliographystyle{ieeetr}
\bibliography{references}

\appendix 

Lemma~\ref{KeyMarkovLemma} below provides upper bounds on the tail
probabilities of sums involving a binary-valued Markov chain. 

\begin{aplemma}\label{KeyMarkovLemma} Let $\{\xi(i)\in\{0,1\}\}_{i\in\mathbb{N}_{0}}$
be a time-inhomogeneous Markov chain with transition probabilities
$p_{q,r}\colon\mathbb{N}_{0}\to[0,1]$, $q,r\in\{0,1\}$. Furthermore,
let $\{\chi(i)\in\{0,1\}\}_{i\in\mathbb{N}_{0}}$ be a binary-valued
process that is independent of $\{\xi(i)\in\{0,1\}\}_{i\in\mathbb{N}_{0}}$.
Assume 
\begin{align}
 & p_{q,1}(i)\leq\tilde{p},\,\,q\in\{0,1\},\,\,i\in\mathbb{N}_{0},\label{eq:xicond}\\
 & \sum_{k=1}^{\infty}\mathbb{P}[\sum_{i=0}^{k-1}\chi(i)>\tilde{w}k]<\infty,\label{eq:chicond}
\end{align}
where $\tilde{p}\in(0,1)$, $\tilde{w}\in(0,1]$. We then have for
$\rho\in(\tilde{p}\tilde{w},\tilde{w})$, 
\begin{align}
\mathbb{P}[\sum_{i=0}^{k-1}\xi(i)\chi(i)>\rho k] & \leq\psi_{k},\quad k\in\mathbb{N},\label{eq:keylemmaresult1}
\end{align}
 where $\psi_{k}\triangleq\tilde{\sigma}_{k}+\phi^{-\rho k+1}\frac{\left((\phi-1)\tilde{p}+1\right)^{\tilde{w}k}-1}{(\phi-1)\tilde{p}}$
with $\phi\triangleq\frac{\frac{\rho}{\tilde{w}}(1-\tilde{p})}{\tilde{p}(1-\frac{\rho}{\tilde{w}})}$,
$\tilde{\sigma}_{k}\triangleq\mathbb{P}[\sum_{i=0}^{k-1}\chi(i)>\tilde{w}k],k\in\mathbb{N}$.
Moreover, $\sum_{k=1}^{\infty}\psi_{k}<\infty.$ 

\end{aplemma}

\vskip 10pt

In the proof of Lemma~\ref{KeyMarkovLemma}, by following the approach
used for obtaining Chernoff-type tail distribution inequalities for
sums of independent random variables (see Appendix B of \cite{madhow2008fundamentals}
and Section 1.9 of \cite{billingsley1986}) we use Markov's inequality.
Specifically, let $y$ denote the sum of a number of random variables,
and consider the tail probability $\mathbb{P}[y>\varsigma]$, where
$\varsigma\in\mathbb{R}$. In obtaining a bound for this tail probability,
Markov's inequality is utilized to obtain 
\begin{align*}
\mathbb{P}[y & >\varsigma]\leq\mathbb{P}[y\geq\varsigma]=\mathbb{P}[\phi^{y}\geq\phi^{\varsigma}]\leq\phi^{-\varsigma}\mathbb{E}[\phi^{y}],
\end{align*}
 for $\phi>1$. Chernoff bound is then given by $\min_{\phi>1}\phi^{-\varsigma}\mathbb{E}[\phi^{y}]$. 

In the proof of Lemma~\ref{KeyMarkovLemma} we do not provide the
details of the minimization process to obtain $\phi$ that gives the
optimum bound. Instead, we show that the tail probability inequality
(\ref{eq:keylemmaresult1}) holds with $\psi_{k},k\in\mathbb{N},$
and that $\sum_{k=1}^{\infty}\psi_{k}<\infty$. To obtain this result,
in addition to Markov's inequality, some additional key steps (including
Lemma~\ref{ExpectationLemma} below) are also required due to the
fact that in Lemma~\ref{KeyMarkovLemma} we consider sums of (not
necessarily independent) random variables composed of the product
of states of a time-inhomogeneous Markov chain and a binary-valued
process that satisfy (\ref{eq:chicond}).

\vskip 10pt

\begin{aplemma} \label{ExpectationLemma} Let $\{\xi(i)\in\{0,1\}\}_{i\in\mathbb{N}_{0}}$
be an $\mathcal{F}_{i}$-adapted binary-valued Markov chain with transition
probability functions $p_{q,r}\colon\mathbb{N}_{0}\to[0,1]$, $q,r\in\{0,1\}$.
Then for all $\phi>1$, $s\in\mathbb{N}$, and $\tilde{p}\in[0,1]$
such that 
\begin{align}
p_{q,1} & (i)\leq\tilde{p},\,\,q\in\{0,1\},\,\,i\in\mathbb{N}_{0},\label{eq:transitioncondition}
\end{align}
 we have 
\begin{align}
\mathbb{E}[\phi^{\sum_{j=1}^{s}\xi(i_{j})}] & \leq\phi\left((\phi-1)\tilde{p}+1\right)^{s-1},\label{eq:lemmaresultfors}
\end{align}
 where $i_{1},i_{2},\ldots,i_{s}\in\mathbb{N}_{0}$ denote indices
such that $0\leq i_{1}<i_{2}<\ldots<i_{s}$. \end{aplemma}

\begin{IEEEproof} We show by induction. First, for the case $s=1$,
\begin{align}
\mathbb{E}[\phi^{\sum_{j=1}^{s}\xi(i_{j})}] & =\mathbb{E}[\phi^{\xi(i_{1})}]\leq\phi.\label{eq:resultfors1}
\end{align}
For the case $s=2$, the random variable $\xi(i_{1})$ is $\mathcal{F}_{i_{2}-1}$-measurable
(because $i_{1}\leq i_{2}-1$), and thus we have 
\begin{align}
\mathbb{E}[\phi^{\sum_{j=1}^{s}\xi(i_{j})}] & =\mathbb{E}[\phi^{\xi(i_{1})}\phi^{\xi(i_{2})}]\nonumber \\
 & =\mathbb{E}[\mathbb{E}[\phi^{\xi(i_{1})}\phi^{\xi(i_{2})}\mid\mathcal{F}_{i_{2}-1}]]\nonumber \\
 & =\mathbb{E}[\phi^{\xi(i_{1})}\mathbb{E}[\phi^{\xi(i_{2})}\mid\mathcal{F}_{i_{2}-1}]].\label{eq:cases2part1}
\end{align}
 Noting that $\{\xi(i)\in\{0,1\}\}_{i\in\mathbb{N}_{0}}$ is a Markov
chain, we obtain $\mathbb{E}[\phi^{\xi(i_{2})}\mid\mathcal{F}_{i_{2}-1}]=\mathbb{E}[\phi^{\xi(i_{2})}\mid\xi(i_{2}-1)]$.
Consequently, 
\begin{align}
 & \mathbb{E}[\phi^{\sum_{j=1}^{s}\xi(i_{j})}]=\mathbb{E}[\phi^{\xi(i_{1})}\mathbb{E}[\phi^{\xi(i_{2})}\mid\xi(i_{2}-1)]]\nonumber \\
 & \,\,=\mathbb{E}\Big[\phi^{\xi(i_{1})}\Big(\phi\mathbb{P}[\xi(i_{2})=1\mid\xi(i_{2}-1)]\nonumber \\
 & \,\,\quad\quad+\mathbb{P}[\xi(i_{2})=0\mid\xi(i_{2}-1)]\Big)\Big]\nonumber \\
 & \,\,=\mathbb{E}\Big[\phi^{\xi(i_{1})}\Big(\phi\mathbb{P}[\xi(i_{2})=1\mid\xi(i_{2}-1)]\nonumber \\
 & \,\,\quad\quad+1-\mathbb{P}[\xi(i_{2})=1\mid\xi(i_{2}-1)]\Big)\Big]\nonumber \\
 & \,\,=\mathbb{E}\Big[\phi^{\xi(i_{1})}\Big((\phi-1)\mathbb{P}[\xi(i_{2})=1\mid\xi(i_{2}-1)]+1\Big)\Big].\label{eq:cases2part2}
\end{align}
Then by using (\ref{eq:transitioncondition}) and (\ref{eq:resultfors1}),
we arrive at 
\begin{align}
 & \mathbb{E}[\phi^{\sum_{j=1}^{s}\xi(i_{j})}]\leq\mathbb{E}\Big[\phi^{\xi(i_{1})}\Big((\phi-1)\tilde{p}+1\Big)\Big]\nonumber \\
 & \quad=\mathbb{E}[\phi^{\xi(i_{1})}]((\phi-1)\tilde{p}+1)\leq\phi((\phi-1)\tilde{p}+1).\label{eq:resultfors2}
\end{align}
Hence, we have that (\ref{eq:lemmaresultfors}) is satisfied for $s\in\{1,2\}$. 

Now, suppose that (\ref{eq:lemmaresultfors}) holds for $s=\tilde{s}>2$,
that is, 
\begin{align}
\mathbb{E}[\phi^{\sum_{j=1}^{\tilde{s}}\xi(i_{j})}] & \leq\phi\left((\phi-1)\tilde{p}+1\right)^{\tilde{s}-1}.\label{eq:resulttildes}
\end{align}
We must show that (\ref{eq:lemmaresultfors}) holds for $s=\tilde{s}+1$.
Using arguments similar to those used for obtaining (\ref{eq:cases2part1})--(\ref{eq:resultfors2}),
we obtain 
\begin{align}
\mathbb{E}[\phi^{\sum_{j=1}^{\tilde{s}+1}\xi(i_{j})}] & =\mathbb{E}[\phi^{\sum_{j=1}^{\tilde{s}}\xi(i_{j})}\phi^{\xi(i_{\tilde{s}+1})}]\nonumber \\
 & =\mathbb{E}[\mathbb{E}[\phi^{\sum_{j=1}^{\tilde{s}}\xi(i_{j})}\phi^{\xi(i_{\tilde{s}+1})}\mid\mathcal{F}_{i_{\tilde{s}+1}-1}]]\nonumber \\
 & =\mathbb{E}[\phi^{\sum_{j=1}^{\tilde{s}}\xi(i_{j})}\mathbb{E}[\phi^{\xi(i_{\tilde{s}+1})}\mid\mathcal{F}_{i_{\tilde{s}+1}-1}]]\nonumber \\
 & =\mathbb{E}[\phi^{\sum_{j=1}^{\tilde{s}}\xi(i_{j})}\mathbb{E}[\phi^{\xi(i_{\tilde{s}+1})}\mid\xi(i_{\tilde{s}+1}-1)]]\nonumber \\
 & \leq\mathbb{E}[\phi^{\sum_{j=1}^{\tilde{s}}\xi(i_{j})}]((\phi-1)\tilde{p}+1).\label{eq:casesfortildesplus1part1}
\end{align}
Using (\ref{eq:resulttildes}) and (\ref{eq:casesfortildesplus1part1}),
we arrive at (\ref{eq:lemmaresultfors}) with $s=\tilde{s}+1$. \end{IEEEproof}

\vskip 10pt

\emph{Proof of Lemma~\ref{KeyMarkovLemma}:} First, let 
\begin{align*}
\overline{\xi}(k) & \triangleq[\xi(0),\xi(1),\ldots,\xi(k-1)]^{\mathrm{T}},\\
\overline{\chi}(k) & \triangleq[\chi(0),\chi(1),\ldots,\chi(k-1)]^{\mathrm{T}},\quad k\in\mathbb{N}.
\end{align*}
 Now let 
\begin{align*}
F_{s,k} & \triangleq\{\overline{\chi}\in\{0,1\}^{k}\colon\overline{\chi}^{\mathrm{T}}\overline{\chi}=s\},\,\,s\in\{0,1,\ldots,k\},\,k\in\mathbb{N}.
\end{align*}
 It is clear that $F_{s_{1},k}\cap F_{s_{2},k}=\emptyset$, $s_{1}\neq s_{2}$;
moreover, 
\begin{align*}
\mathbb{P}[\overline{\chi}(k)\in\cup_{s=0}^{k}F_{s,k}] & =1,\quad k\in\mathbb{N}.
\end{align*}
It then follows that for all $\rho\in(\tilde{p}\tilde{w},1)$ and
$k\in\mathbb{N}$, 
\begin{align}
 & \mathbb{P}[\sum_{i=0}^{k-1}\xi(i)\chi(i)>\rho k]=\mathbb{P}[\overline{\xi}^{\mathrm{T}}(k)\overline{\chi}(k)>\rho k]\nonumber \\
 & \quad=\sum_{s=0}^{k}\sum_{\overline{\chi}\in F_{s,k}}\mathbb{P}[\overline{\xi}^{\mathrm{T}}(k)\overline{\chi}(k)>\rho k\mid\overline{\chi}(k)=\overline{\chi}]\nonumber \\
 & \quad\quad\cdot\mathbb{P}[\overline{\chi}(k)=\overline{\chi}].\label{eq:xibarchibarequation}
\end{align}
Due to the mutual independence of $\xi(\cdot)$ and $\chi(\cdot)$,
\begin{align}
\mathbb{P}[\overline{\xi}^{\mathrm{T}}(k)\overline{\chi}(k)>\rho k\mid\overline{\chi}(k)=\overline{\chi}] & =\mathbb{P}[\overline{\xi}^{\mathrm{T}}(k)\overline{\chi}>\rho k].\label{eq:conditionalprobabilityresolution}
\end{align}
As a result, it follows from (\ref{eq:xibarchibarequation}) and (\ref{eq:conditionalprobabilityresolution})
that for $k\in\mathbb{N}$, 
\begin{align}
 & \mathbb{P}[\sum_{i=0}^{k-1}\xi(i)\chi(i)>\rho k]\nonumber \\
 & \,\,=\sum_{s=0}^{k}\sum_{\overline{\chi}\in F_{s,k}}\mathbb{P}[\overline{\xi}^{\mathrm{T}}(k)\overline{\chi}>\rho k]\mathbb{P}[\overline{\chi}(k)=\overline{\chi}]\nonumber \\
 & \,\,=\sum_{s=0}^{\lfloor\tilde{w}k\rfloor}\sum_{\overline{\chi}\in F_{s,k}}\mathbb{P}[\overline{\xi}^{\mathrm{T}}(k)\overline{\chi}>\rho k]\mathbb{P}[\overline{\chi}(k)=\overline{\chi}]\nonumber \\
 & \,\,\quad+\sum_{s=\lfloor\tilde{w}k\rfloor+1}^{k}\sum_{\overline{\chi}\in F_{s,k}}\mathbb{P}[\overline{\xi}^{\mathrm{T}}(k)\overline{\chi}>\rho k]\mathbb{P}[\overline{\chi}(k)=\overline{\chi}].\label{eq:new-two-terms}
\end{align}

In the following, we will find upper-bounds for the two summation
terms in (\ref{eq:new-two-terms}). First, for the second term, since
$\mathbb{P}[\overline{\xi}^{\mathrm{T}}(k)\overline{\chi}>\rho k]\leq1$,
$k\in\mathbb{N}$, we have 
\begin{align}
 & \sum_{s=\lfloor\tilde{w}k\rfloor+1}^{k}\sum_{\overline{\chi}\in F_{s,k}}\mathbb{P}[\overline{\xi}^{\mathrm{T}}(k)\overline{\chi}>\rho k]\mathbb{P}[\overline{\chi}(k)=\overline{\chi}]\nonumber \\
 & \quad\leq\sum_{s=\lfloor\tilde{w}k\rfloor+1}^{k}\sum_{\overline{\chi}\in F_{s,k}}\mathbb{P}[\overline{\chi}(k)=\overline{\chi}]\nonumber \\
 & \quad=\mathbb{P}[\sum_{i=0}^{k-1}\chi(i)>\tilde{w}k]=\tilde{\sigma}_{k},\quad k\in\mathbb{N}.\label{eq:finalsigmainequality}
\end{align}
Next, we look at the first term in (\ref{eq:new-two-terms}). Note
that $\mathbb{P}[\overline{\xi}^{\mathrm{T}}(k)\overline{\chi}>\rho k]=0$
for $\overline{\chi}\in F_{0,k}$. Hence, for all $k\in\mathbb{N}$
such that $\lfloor\tilde{w}k\rfloor=0$, we have 
\begin{align}
\sum_{s=0}^{\lfloor\tilde{w}k\rfloor}\sum_{\overline{\chi}\in F_{s,k}}\mathbb{P}[\overline{\xi}^{\mathrm{T}}(k)\overline{\chi}>\rho k]\mathbb{P}[\overline{\chi}(k)=\overline{\chi}] & =0.
\end{align}
 Furthermore, for all $k\in\mathbb{N}$ such that $\lfloor\tilde{w}k\rfloor\geq1$,
we have 
\begin{align}
 & \sum_{s=0}^{\lfloor\tilde{w}k\rfloor}\sum_{\overline{\chi}\in F_{s,k}}\mathbb{P}[\overline{\xi}^{\mathrm{T}}(k)\overline{\chi}>\rho k]\mathbb{P}[\overline{\chi}(k)=\overline{\chi}]\nonumber \\
 & \,=\sum_{s=1}^{\lfloor\tilde{w}k\rfloor}\sum_{\overline{\chi}\in F_{s,k}}\mathbb{P}[\overline{\xi}^{\mathrm{T}}(k)\overline{\chi}>\rho k]\mathbb{P}[\overline{\chi}(k)=\overline{\chi}].\label{eq:sumoverchibar}
\end{align}
Now, for $s\in\{1,2,\ldots,\lfloor\tilde{w}k\rfloor\}$, let $i_{1}(\overline{\chi}),i_{2}(\overline{\chi}),\ldots,i_{s}(\overline{\chi})$
denote the indices of the nonzero entries of $\overline{\chi}\in F_{s,k}$
such that $i_{1}(\overline{\chi})<i_{2}(\overline{\chi})<\cdots<i_{s}(\overline{\chi})$.
Consequently, 
\begin{align}
\mathbb{P}[\overline{\xi}^{\mathrm{T}}(k)\overline{\chi}>\rho k] & =\mathbb{P}[\sum_{j=1}^{s}\overline{\xi}_{i_{j}(\bar{\chi})}(k)>\rho k]\nonumber \\
 & =\mathbb{P}[\sum_{j=1}^{s}\xi(i_{j}(\overline{\chi})-1)>\rho k],\label{eq:xibarequation}
\end{align}
for $\overline{\chi}\in F_{s,k}$, $s\in\{1,2,\ldots,\lfloor\tilde{w}k\rfloor\}$,
and $k\in\mathbb{N}$ such that $\lfloor\tilde{w}k\rfloor\geq1$. 

Now note that $\phi>1$, since $\rho\in(\tilde{p}\tilde{w},\tilde{w})$.
We use Markov's inequality to obtain 
\begin{align}
\mathbb{P}[\overline{\xi}^{\mathrm{T}}(k)\overline{\chi}>\rho k] & \leq\mathbb{P}[\sum_{j=1}^{s}\xi(i_{j}(\overline{\chi})-1)\geq\rho k]\nonumber \\
 & =\mathbb{P}[\phi^{\sum_{j=1}^{s}\xi(i_{j}(\overline{\chi})-1)}\geq\phi^{\rho k}]\nonumber \\
 & \leq\phi^{-\rho k}\mathbb{E}[\phi^{\sum_{j=1}^{s}\xi(i_{j}(\overline{\chi})-1)}].\label{eq:xiphiineq}
\end{align}
It follows from Lemma~\ref{ExpectationLemma} that $\mathbb{E}[\phi^{\sum_{j=1}^{s}\xi(i_{j}(\overline{\chi})-1)}]\leq\phi\left((\phi-1)\tilde{p}+1\right)^{s-1}$.
Using this inequality together with (\ref{eq:sumoverchibar}) and
(\ref{eq:xiphiineq}), for all $k\in\mathbb{N}$ such that $\lfloor\tilde{w}k\rfloor\geq1$,
we obtain 
\begin{align}
 & \sum_{s=0}^{\lfloor\tilde{w}k\rfloor}\sum_{\overline{\chi}\in F_{s,k}}\mathbb{P}[\overline{\xi}^{\mathrm{T}}(k)\overline{\chi}>\rho k]\mathbb{P}[\overline{\chi}(k)=\overline{\chi}]\nonumber \\
 & \,\leq\sum_{s=1}^{\lfloor\tilde{w}k\rfloor}\sum_{\overline{\chi}\in F_{s,k}}\phi^{-\rho k}\phi\left((\phi-1)\tilde{p}+1\right)^{s-1}\mathbb{P}[\overline{\chi}_{k}=\overline{\chi}]\nonumber \\
 & \,=\phi^{-\rho k+1}\sum_{s=1}^{\lfloor\tilde{w}k\rfloor}\left((\phi-1)\tilde{p}+1\right)^{s-1}\sum_{\overline{\chi}\in F_{s,k}}\mathbb{P}[\overline{\chi}_{k}=\overline{\chi}]\nonumber \\
 & \,=\phi^{-\rho k+1}\sum_{s=1}^{\lfloor\tilde{w}k\rfloor}\left((\phi-1)\tilde{p}+1\right)^{s-1}\mathbb{P}[\overline{\chi}_{k}\in F_{s,k}]\nonumber \\
 & \,\leq\phi^{-\rho k+1}\sum_{s=1}^{\lfloor\tilde{w}k\rfloor}\left((\phi-1)\tilde{p}+1\right)^{s-1},\label{eq:finaxichiinequality}
\end{align}
where we also used the fact that $\mathbb{P}[\overline{\chi}_{k}\in F_{s,k}]\leq1$
to obtain the last inequality. Here, we have 
\begin{align}
\sum_{s=1}^{\lfloor\tilde{w}k\rfloor}\left((\phi-1)\tilde{p}+1\right)^{s-1} & =\frac{\left((\phi-1)\tilde{p}+1\right)^{\lfloor\tilde{w}k\rfloor}-1}{\left((\phi-1)\tilde{p}+1\right)-1}\nonumber \\
 & \leq\frac{\left((\phi-1)\tilde{p}+1\right)^{\tilde{w}k}-1}{(\phi-1)\tilde{p}}.\label{eq:geometricseriesfinitesum}
\end{align}
Hence, (\ref{eq:finaxichiinequality}) and (\ref{eq:geometricseriesfinitesum})
imply 
\begin{align}
 & \sum_{s=0}^{\lfloor\tilde{w}k\rfloor}\sum_{\overline{\chi}\in F_{s,k}}\mathbb{P}[\overline{\xi}^{\mathrm{T}}(k)\overline{\chi}>\rho k]\mathbb{P}[\overline{\chi}(k)=\overline{\chi}]\nonumber \\
 & \quad\leq\phi^{-\rho k+1}\frac{\left((\phi-1)\tilde{p}+1\right)^{\tilde{w}k}-1}{(\phi-1)\tilde{p}},\label{eq:finalphiwinequality}
\end{align}
for all $k\in\mathbb{N}$ such that $\lfloor\tilde{w}k\rfloor\geq1$.
Because the right-hand side of this inequality is zero if $\lfloor\tilde{w}k\rfloor=0$,
(\ref{eq:finalphiwinequality}) holds for all $k\in\mathbb{N}$. Now,
this fact together with (\ref{eq:new-two-terms}), (\ref{eq:finalsigmainequality})
leads us to (\ref{eq:keylemmaresult1}).

Our next goal is to show $\sum_{k=1}^{\infty}\psi_{k}<\infty$. To
this end, first note that 
\begin{align}
 & \sum_{k=1}^{\infty}\phi^{-\rho k+1}\frac{\left((\phi-1)\tilde{p}+1\right)^{\tilde{w}k}-1}{(\phi-1)\tilde{p}}\nonumber \\
 & \quad=\frac{\phi}{(\phi-1)\tilde{p}}\sum_{k=1}^{\infty}\phi^{-\rho k}\left((\phi-1)\tilde{p}+1\right)^{\tilde{w}k}\nonumber \\
 & \quad\quad-\frac{\phi}{(\phi-1)\tilde{p}}\sum_{k=1}^{\infty}\phi^{-\rho k}.\label{eq:psisum}
\end{align}
We will show that the series on the far right-hand side of (\ref{eq:psisum})
are both convergent. First, since $\phi>1$, we have $\phi^{-\rho}<1$,
and thus, the geometric series $\sum_{k=1}^{\infty}\phi^{-\rho k}$
converges, that is, 
\begin{align}
\sum_{k=1}^{\infty}\phi^{-\rho k} & <\infty.\label{eq:easyconvergentsum}
\end{align}
Next, we show $\phi^{-\rho}\left((\phi-1)\tilde{p}+1\right)^{\tilde{w}}<1$.
We obtain 
\begin{align}
\phi^{-\rho}\left((\phi-1)\tilde{p}+1\right)^{\tilde{w}} & =\left(\phi^{-\frac{\rho}{\tilde{w}}}\left((\phi-1)\tilde{p}+1\right)\right)^{\tilde{w}}.\label{eq:wpower}
\end{align}
 Furthermore, 
\begin{align*}
 & \phi^{-\frac{\rho}{\tilde{w}}}\left((\phi-1)\tilde{p}+1\right)\\
 & \quad=\left(\frac{\frac{\rho}{\tilde{w}}(1-\tilde{p})}{\tilde{p}(1-\frac{\rho}{\tilde{w}})}\right)^{-\frac{\rho}{\tilde{w}}}\left(\left(\frac{\frac{\rho}{\tilde{w}}(1-\tilde{p})}{\tilde{p}(1-\frac{\rho}{\tilde{w}})}-1\right)\tilde{p}+1\right)\\
 & \quad=\left(\frac{\tilde{p}\tilde{w}}{\rho}\right)^{\frac{\rho}{\tilde{w}}}\left(\frac{1-\tilde{p}}{1-\frac{\rho}{\tilde{w}}}\right)^{-\frac{\rho}{\tilde{w}}}\left(\frac{1-\tilde{p}}{1-\frac{\rho}{\tilde{w}}}\right)\\
 & \quad=\left(\frac{\tilde{p}\tilde{w}}{\rho}\right)^{\frac{\rho}{\tilde{w}}}\left(\frac{1-\tilde{p}}{1-\frac{\rho}{\tilde{w}}}\right)^{1-\frac{\rho}{\tilde{w}}}.
\end{align*}
 Note that $\frac{\tilde{p}\tilde{w}}{\rho},\frac{1-\tilde{p}}{1-\frac{\rho}{\tilde{w}}}\in(0,1)\cup(1,\infty)$.
Since $\ln v<v-1$ for any $v\in(0,1)\cup(1,\infty)$, we have 
\begin{align*}
 & \ln\left(\phi^{-\frac{\rho}{\tilde{w}}}\left((\phi-1)\tilde{p}+1\right)\right)\\
 & \quad=\frac{\rho}{\tilde{w}}\ln\left(\frac{\tilde{p}\tilde{w}}{\rho}\right)+(1-\frac{\rho}{\tilde{w}})\ln\left(\frac{1-\tilde{p}}{1-\frac{\rho}{\tilde{w}}}\right)\\
 & \quad<\frac{\rho}{\tilde{w}}\left(\frac{\tilde{p}\tilde{w}}{\rho}-1\right)+(1-\frac{\rho}{\tilde{w}})\left(\frac{1-\tilde{p}}{1-\frac{\rho}{\tilde{w}}}-1\right)\\
 & \quad=\tilde{p}-\frac{\rho}{\tilde{w}}+\frac{p}{\tilde{w}}-\tilde{p}\,=\,0,
\end{align*}
 which implies that $\phi^{-\frac{\rho}{\tilde{w}}}\left((\phi-1)\tilde{p}+1\right)<1$,
and hence by (\ref{eq:wpower}), $\phi^{-\rho}\left((\phi-1)\tilde{p}+1\right)^{\tilde{w}}<1$.
Therefore, 
\begin{align}
\sum_{k=1}^{\infty}\phi^{-\rho k}\left((\phi-1)\tilde{p}+1\right)^{\tilde{w}k}<\infty.\label{eq:difficultcasesum}
\end{align}
Finally, (\ref{eq:psisum}), (\ref{eq:easyconvergentsum}), and (\ref{eq:difficultcasesum})
imply $\sum_{k=1}^{\infty}\psi_{k}<\infty$. $\hfill \square$

\end{document}